\documentclass[aps,prx,reprint,twocolumn,superscriptaddress,nofootinbib]{revtex4-2}

\usepackage{graphicx}
\usepackage{bm}
\usepackage{amsmath}
\usepackage{amssymb}
\usepackage{amsfonts}
\usepackage{amsthm}
\usepackage{mathtools,faktor}
\usepackage{hyperref}
\usepackage[dvipsnames]{xcolor}
\usepackage{comment}

\def\h{\mathbf h}
\def\v{\mathbf v}
\def\u{\mathbf u}
\def\W{\mathbf W}
\def\z{\mathbf z}

\renewcommand{\Re}{\operatorname{Re}}

\hypersetup{colorlinks=true,urlcolor=[rgb]{0,0,0.5},citecolor=[rgb]{0.5,0,0},linkcolor=[rgb]{0,0,0.4}}

\begin{document}
\title{Dynamics of learning to integrate in linear recurrent neural networks}

\author{Blake Bordelon}
\email{blake\_bordelon@g.harvard.edu}
\affiliation{John A. Paulson School of Engineering and Applied Sciences, Harvard University, Cambridge, MA, USA}
\affiliation{Center for Brain Science, Harvard University, Cambridge, MA, USA}

\author{Jordan Cotler}
\email{jcotler@fas.harvard.edu}
\affiliation{Department of Physics, Harvard University, Cambridge, MA, USA}

\author{Cengiz Pehlevan}
\email{cpehlevan@seas.harvard.edu}
\affiliation{John A. Paulson School of Engineering and Applied Sciences, Harvard University, Cambridge, MA, USA}
\affiliation{Center for Brain Science, Harvard University, Cambridge, MA, USA}
\affiliation{Kempner Institute for the Study of Natural and Artificial Intelligence, Harvard University, Cambridge, MA, USA}

\author{Jacob A. Zavatone-Veth}
\email{jzavatoneveth@fas.harvard.edu}
\affiliation{Center for Brain Science, Harvard University, Cambridge, MA, USA}
\affiliation{Society of Fellows, Harvard University, Cambridge, MA, USA}




\begin{abstract}
    Learning recurrent connectivity that supports memory over long intrinsic timescales is a basic problem in the theory of dynamical computation. While continuous attractor and integrator models describe how tuned recurrent circuits can maintain information, less is known about how such slow modes are acquired by gradient-based learning. Here we study this question in an analytically tractable setting: we build a mathematical theory of the learning dynamics of linear RNNs trained to integrate white noise. We show that when the initial recurrent weights are small, the dynamics of learning are described by a low-dimensional system that tracks a single outlier eigenvalue of the recurrent weights. This reveals the precise manner in which the long timescale associated with white noise integration is learned.  We extend our analyses to RNNs learning a damped oscillatory filter, and find low-dimensional effective dynamical equations for the evolution of a conjugate pair of outlier eigenvalues.  Taken together, our analyses build a rich mathematical framework for studying dynamical learning problems relevant to both machine learning and neuroscience. 
\end{abstract}

\maketitle

\section{Introduction}

Recurrent neural networks (RNNs) are a canonical model of dynamical computation, both as simplified models for biological neural circuits and as sequence models in machine learning~\cite{rosenblatt1962principles,hopfield1982neural}. In neuroscience, RNNs with connectivity structured to implement continuous attractor or integrator dynamics are believed to underlie many forms of working memory in the brain~\cite{khona2022attractor}. In machine learning, vanilla RNNs have historically provided a core model class for sequence processing~\cite{hochreiter1997lstm,orvieto2023resurrecting,gu2020hippo}. 

Understanding dynamical memory in RNNs requires both an understanding of what circuit mechanisms can sustain long-timescale memories, and of how the required connectivity can be learned. If connections are tuned and stable, a variety of connectivity motifs can sustain slow modes without ongoing learning signals \cite{goldman2009feedback,khona2022attractor,seung1996still,ganguli2008traces}. If connections must be learned \textit{ab initio} from data using gradient descent, optimization dynamics in vanilla RNNs can be delicate~\cite{hochreiter1991untersuchungen,hochreiter1997lstm,orvieto2023resurrecting,zucchet2024recurrent,li2021curse,cohen2023learning,engelken2023flossing}. This challenge has motivated the development of alternative architectures with more favorable optimization properties, which are central to modern sequence modeling~\cite{gu2020hippo,orvieto2023resurrecting,hochreiter1997lstm,gu2024mamba,vaswani2017attention}. Nonetheless, even in simple recurrent models, we lack a detailed theoretical account of how gradient descent creates the slow dynamical modes required for temporal integration. 

This question is analogous to a central theme in the theory of feedforward neural networks: the goal of understanding how gradient descent dynamically selects among possible solutions to a given task. Much progress in the feedforward setting has been made by distinguishing between a lazy regime where parameters move infinitesimally and learning dynamics are linear, and a rich regime where features are learned~\cite{chizat2019lazy,woodworth2020kernel,yang2021feature,bordelon2022dmft}. Though the nonlinear dynamics of learning in the rich regime are generally intractable, substantial insight can be gained from studying deep \textit{linear} feedforward networks~\cite{fukumizu1998effect,saxe2013exact,atanasov2022alignment,braun2022exact,bordelon2025deep}. Deep linear networks are trivial from the perspective of which functions they can express, but depth leads to richly nonlinear learning dynamics. These analyses reveal a two-stage learning process in which the weights first align to task-relevant directions and then grow in scale. In the recurrent setting, while a lazy regime is known~\cite{alemohammad2020recurrent} and past works have uncovered signatures of alignment of low-rank components of recurrent weights~\cite{schuessler2024oblique,maheswaranathan2019reverse,schuessler2020interplay,liu2024how,cohen2023learning,haputhanthri2025understanding,mastrogiuseppe2018linking,schuessler2020interplay,schuessler2024oblique}, we still lack a detailed analytical understanding in simple settings akin to the work on feedforward networks. 

To characterize how gradient descent creates slow recurrent timescales in a controlled setting, we build a mathematical theory of the learning dynamics of linear RNNs trained to integrate white noise. We show that when the initial recurrent weights are small, the dynamics of learning are described by a low-dimensional system that tracks a single outlier eigenvalue of the recurrent weights. This characterizes how the long timescale associated with white noise integration is learned. We extend our analyses to RNNs learning a damped oscillatory filter, and find low-dimensional effective dynamics for the evolution of a conjugate pair of outlier eigenvalues. Our results provide a mathematical framework for studying how gradient descent sculpts recurrent dynamics.

\begin{figure*}[t!]
    \centering
    \includegraphics[width=6.74in]{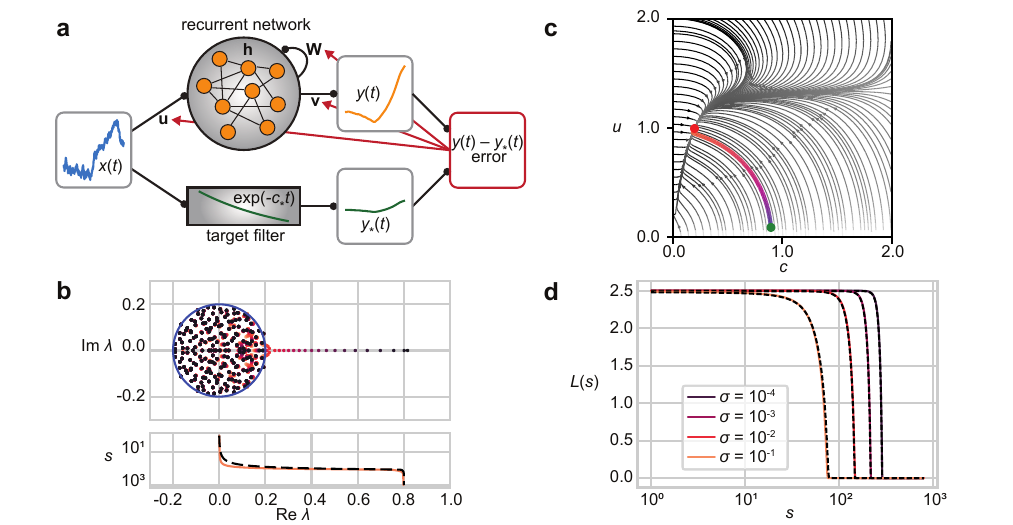}
    \caption{\textbf{Training a linear RNN to integrate noise.} \textbf{a}. Diagram of training setup. Gaussian noise $x(t)$ is fed into a linear RNN with recurrent weights $\W$ through a read-in vector $\u$, and the resulting signal $y(t)$ is read out through a vector $\v$. This signal is then compared to a target signal $y_{\star}(t)$ generated by filtering the noise signal with an exponential filter $e^{-c_{\star} t}$, and the error signal is backpropagated to update the parameters (diagrammed as red arrows).  \textbf{b}. 
    Dynamics for $c_\star = 0.2$ and $\sigma = 0.2$. (Top) Eigenvalues of $\frac{1}{\sqrt{N}} \W(s)$ during training of a $N=250$ network. Shading indicates training steps, with darker colors indicating later times. (Bottom) Comparison of evolution of the projection of the recurrent weights onto the read-in/out directions $\frac{\v^{\top} \W \u}{\Vert \v \Vert\, \Vert \u \Vert}$, which measures the outlier eigenvalue after alignment, (orange line) compared to the theoretical prediction $1-c(s)$ from Equation \eqref{E:maindyn1} (black dashed line). \textbf{c}. Flow field of the reduced two-dimensional gradient flow dynamics for $c_{\star} = 0.2$, with the trajectory for $\sigma = 0.2$ overlaid. The green dot indicates the initial condition for the reduced dynamics at the end of the alignment phase, and the red dot indicates the endpoint. Colors along the path indicate training time. \textbf{d}. Loss dynamics for varying initialization scale $\sigma$. The initial alignment takes time $t \propto - \ln(\sigma)$. 
    }
    \label{fig:real_target}
\end{figure*}

\section{Setup} 

Our setup comprises a streamlined version of previous works on training RNNs to solve neuroscience and machine learning tasks~\cite{schuessler2024oblique,schuessler2020interplay,maheswaranathan2019reverse,maheswaranathan2019individuality,cohen2023learning}. This simplified architecture allows us to obtain a precise analytical characterization of learning dynamics, shedding light on the mechanisms by which RNNs can learn to represent long timescales.

We consider learning in linear RNNs in continuous time~\cite{sompolinsky1988chaos}, with dynamics
\begin{equation}
\begin{split}
    (\partial_{t}+1) \h (t) &= \frac{1}{\sqrt N} \W \h(t) + \u \,x(t), \\ y(t) &= \frac{1}{\gamma N} \v^{\top} \h (t)\,.
\end{split}
\end{equation}
Here there are $N$ neurons driven by scalar Gaussian noise $x(t)$, recurrent weights $\W$, read-in weights $\u$, and read-out weights $\v$. The learning goal is for the read-out $y(t)$ to match a target signal
\begin{align}
    y_{\star}(t) = \int_{0}^{t} dt'\, f_{\star}(t') x(t-t')
\end{align}
generated by convolving the input signal with a fixed target filter $f_{\ast}(t)$. This task is diagrammed in Figure~\ref{fig:real_target}a. 

Learning is accomplished via gradient flow on the population risk
\begin{align}
\begin{split}
    L &= \lim_{T \to \infty} \mathbb{E}_{x} \frac{1}{T}\int_{0}^{T} dt\, [y(t)-y_{\star}(t)]^{2} 
    \\
    &= \int_{0}^{\infty} dt\, [f(t)-f_{\star}(t)]^2,
\end{split}
\end{align}
where 
\begin{equation}
    f(t) = \frac{1}{\gamma N} \v^{\top} e^{-(\mathbf{I}-\frac{1}{\sqrt{N}} \W)t} \u
\end{equation}
is the filter induced by the linear RNN. 

We initialize the RNN parameters according to an isotropic Gaussian: 
\begin{align}
\begin{split}
    W_{ij} &\sim \mathcal{N}(0, \sigma^2), 
    \\
    u_{i} &\sim \mathcal{N}(0, \sigma^2 ),
    \\
    v_{i} &\sim \mathcal{N}(0, \sigma^2 ) .
\end{split}
\end{align}
Then we update the parameters ($\u$, $\v$, and $\W$) using gradient flow with training time $s$. We will subscript the training time, and often suppress it. Thus, $\W_{s}$ refers to the value of the recurrent weights at training time $s$. Using the convention that the gradients are re-scaled by an overall factor of $1/2$ to cancel the factor of 2 arising from the derivative of the squared error, the gradient flow dynamics are
\begin{equation}
\begin{split}
    \frac{d \W_{s}}{ds} &=  - \frac{\eta_{\W}}{2} \nabla_{\W}\! L = - \eta_{\W} \int_{0}^{\infty} dt\, [f(t) - f_{\star}(t)]  \nabla_{\W}\! f(t), 
    \\
    \frac{d\u_{s}}{ds}  &= - \frac{\eta_{\u}}{2} \nabla_{\u}L = - \eta_{\u} \int_{0}^{\infty} dt\, [f(t) - f_{\star}(t)]  \nabla_{\u} f(t), 
    \\
    \frac{d \v_{s} }{ds} &= - \frac{\eta_{\v}}{2} \nabla_{\v}L = - \eta_{\v} \int_{0}^{\infty} dt\, [f(t) - f_{\star}(t)]  \nabla_{\v} f(t) , 
\end{split}
\end{equation}
where we allow for parameter-specific learning rates $\eta_{\W}$, $\eta_{\u}$, and $\eta_{\v}$. In what follows, we take
\begin{align}
    \eta_{\W}, \eta_\u , \eta_\v = \gamma^2 N
\end{align}
to induce learning on a timescale that is $\mathcal{O}_N(1)$. This is known in the machine learning literature as mean field scaling \cite{bordelon2022dmft,chizat2019lazy}. With this scaling, at large $N$ the filter $f(t)$ starts near zero, and then grows through training. 

We can compute the gradients for $\u_s$,$\v_s$,and $\W_s$ in terms of adjoint variables $\u_{s}(t)$ and $\v_{s}(t)$, which solve
\begin{align} \label{eqn:adjoint_dynamics}
    \frac{d}{dt} \v_{s}(t) &= - \v_{s}(t) + \frac{1}{\sqrt{N}} \W_{s}^{\top} \v_{s}(t) , \quad \v_{s}(0) = \v_{s}
    \\
    \frac{d}{dt} \u_{s}(t) &= - \u_{s}(t) + \frac{1}{\sqrt{N}} \W_{s} \u_{s}(t) , \quad \u(0)_{s} = \u_{s} . 
\end{align}
Then, we have the compact expressions for the gradients:
\begin{align}
\begin{split}
    \nabla_{\u_{s}} f_{s}(t) &= \frac{1}{\gamma N} \v_{s}(t) 
    \\ 
    \nabla_{\v_{s}} f_{s}(t) &= \frac{1}{\gamma N} \u_{s}(t) 
    \\
    \nabla_{\W_{s}} f_{s}(t) &= \frac{1}{\gamma N^{3/2}} \int_{0}^{t} dt'\, \v_{s}(t') \u_{s}(t-t')^{\top}. 
\end{split}
\end{align}
This leads to the dynamics
\begin{align} \label{eq:gf}
\begin{split}
    \frac{d}{ds} \W_{s} &= - \frac{\gamma}{\sqrt N} \int_{0}^{\infty} dt\, [f_{s}(t) - f_{\star}(t)] 
    \\
    &\qquad\qquad\qquad \times \int_{0}^{t} dt'\, \v_{s}(t') \u_{s}(t-t')^{\top}
    \\
    \frac{d}{ds} \u_{s} &= - \gamma \int_{0}^{\infty} dt\, [f_{s}(t) - f_{\star}(t)] \v_{s}(t) 
    \\
    \frac{d}{ds} \v_{s} &= - \gamma \int_{0}^{\infty} dt\, [f_{s}(t) - f_{\star}(t)] \u_{s}(t) .
\end{split}
\end{align}

\section{Learning a leaky integrator}

We first study how the RNN learns to solve the simplest memory task \cite{seung1996still}: damped integration with 
\begin{align}
    f_{\star}(t) = e^{-c_{\star} t}. 
\end{align}
This target filter corresponds to a leaky integrator that accumulates input over time but forgets it exponentially with rate $c_{\star}$. The goal is for the network to emulate this temporal filtering behavior, effectively maintaining a memory of past inputs with an exponential decay.

\subsection{Dynamics from small initial weights: \texorpdfstring{\\}{} rich learning}

When the initial weights are small ($\sigma \ll 1$), the early-time learning dynamics drive alignment of the read-in and read-out vectors, and there emerges an aligned spike in the recurrent weights. Intuitively, the network starts with negligible interactions, and learning first focuses on strengthening the pathways that are most effective for the learning task \cite{atanasov2022alignment}. Once this aligned spike emerges, one can reduce the problem to a set of low-dimensional effective learning dynamics along the direction defined by the spike.

Early in training, all parameters remain of order $\sigma$ and the residual error is dominated by the target filter $f_\star(t) - f_s(t) \approx f_\star(t)$. Then, we can approximate the learning dynamics as
\begin{align}
\begin{split}
    \frac{d}{ds} \W_{s} &\approx  \frac{\gamma}{\sqrt N} \int_{0}^{\infty} dt\, f_{\star}(t) \int_{0}^{t} dt'\, \v_s(t') \u_s(t-t')^{\top}
    \\
    \frac{d}{ds} \u_s &\approx \gamma \int_{0}^{\infty} dt\, f_{\star}(t) \v_s(t) 
    \\
    \frac{d}{ds} \v_s &\approx  \gamma \int_{0}^{\infty} dt\, f_{\star}(t) \u_s(t) .
\end{split}
\end{align}
These early dynamics result in the alignment of $\u$ and $\v$, meaning that the network predominantly processes information along a single direction in the high-dimensional state space. Concretely, by approximating \eqref{eq:gf} to leading order in the initialization scale, we find the ODEs 
\begin{align}
\begin{split}
\frac{d\v_s}{ds} &\approx \frac{\gamma }{c_\star + 1} \,\u_s \,,\quad \frac{d\u_s}{ds} \approx \frac{\gamma }{c_\star + 1} \,\v_s \\
\frac{d\W_s}{ds} &\approx \frac{1}{\sqrt N} \frac{\gamma }{(c_\star+1)^2}\, \v_s \u_s^\top\,.
\end{split}
\end{align}
Letting $a = \frac{\gamma}{1+c_\star}$ represent the initial growth rate of our parameters, we then have 
\begin{align}
    \u_s \approx \frac{1}{2} e^{a s} (\u_0 + \v_0) + \frac{1}{2} e^{-a s} (\u_0 - \v_0)
    \\
    \v_s \approx \frac{1}{2} e^{a s} (\u_0 + \v_0) - \frac{1}{2} e^{-a s}(\u_0 - \v_0)
\end{align}
We see that $\u_s$ and $\v_s$ are quickly aligning to the average of their initial directions: as one grows, it stimulates the growth of the other and they eventually converge to the same vector. Letting ${\u}_+ = \frac{1}{2} (\u_0 + \v_0)$ represent the average initial condition between read-in and read-out which has norm $|\u_+|^2 \simeq \frac{N \sigma^2}{2}$ for large $N$. We introduce a scale-normalized vector $\hat{\u} = \sqrt{ \frac{N}{{|\u_+|^2}}} \ \u_+$ which has $\frac{1}{N}|\hat \u|^2 = 1$. In terms of this vector, the early dynamics of $\W_s$ are approximately growing in the rank-one direction $\hat \u \hat \u^\top$  
\begin{align}
    \frac{1}{\sqrt N} \W_s \approx  \frac{1}{\sqrt N} \W_0 + \frac{a \sigma^2}{4 \gamma N} (e^{2 a s} -1 ) \hat{\u} \hat{\u}^\top. 
\end{align}
This indicates that the recurrent weights $\W$ develop a rank-one structure proportional to $\hat{\u} \hat{\u}^\top$, effectively enhancing the connectivity along the aligned direction. This outlier eigenvalue escapes from the bulk (exceeds $\sigma$) at a timescale 
\begin{align}
    s_{\text{escape}} \approx \frac{1+c_\star}{2 \gamma} \ln\left( 1 + \frac{4 (1+c_\star)}{ \sigma} \right) . 
\end{align}
We illustrate this scaling in Figure~\ref{fig:escape_vs_sigma_supp}, and the consequent fact that the escape time required to reduce the loss decreases with $\sigma$ in Figure \ref{fig:real_target}d. 

After this alignment timescale, the learned filter $f_s(t)$ can be approximated as
\begin{align}
    f_s(t) &= \frac{1}{N \gamma}  \v_s^\top \exp\left( - \left[ \mathbf I - \frac{1}{\sqrt N} \W_s \right] t \right) \u_s  \nonumber
    \\
    &\approx \frac{1}{\gamma} u(s) v(s) e^{- c(s) t}
\end{align}
where we introduced the scalar projections onto the $\u_+$ direction
\begin{align}
    &u(s) \equiv \frac{1}{N } \u_s \cdot \hat{\u} \ , \ v(s) \equiv \frac{1}{N} \v_s \cdot \hat{\u}  \nonumber
    \\
    &c(s)\equiv 1 -  \frac{1}{N} \hat{\u}^\top \left( \frac{1}{\sqrt N} \W_s \hat{\u} \right). 
\end{align}

Thus, after the early period where the alignment to $\hat{\u}$ direction takes place, the gradients become confined to the $\hat{\u}$ direction:
\begin{align}
    \nabla_{\u} L \propto \hat{\u}, \ \nabla_{\v} L \propto \hat{\u}, \ \nabla_{\W} L \propto \hat{\u} \hat{\u}^\top.
\end{align}
This allows us to describe the learning dynamics entirely in terms of the variables $u(s), v(s), c(s)$. By projecting the full gradient descent updates onto $\hat{\u}$ and using the approximate form of $f_{s}(t)$, we find that these variables evolve according to the effective dynamics
\begin{align}
\begin{split}
\frac{dc(s)}{ds} &= u(s) v(s) \left( \frac{u(s) v(s)}{2 c(s)^2} - \frac{2}{(c(s)+c_\star)^2} \right) \\
\frac{du(s)}{ds} &= - v(s) \left( \frac{u(s) v(s)}{c(s)} - \frac{2}{c(s)+c_\star} \right) \\
\frac{dv(s)}{ds} &= - u(s) \left( \frac{u(s) v(s)}{c(s)} - \frac{2}{c(s)+c_\star} \right),
\end{split}
\end{align}
where from here on we set $\gamma = 1$ for simplicity.

We can further simplify the effective dynamics by noting that they have a conservation law
\begin{align}
    \frac{d}{ds} u(s)^2 = \frac{d}{ds} v(s)^2,
\end{align}
so if $v(0) = u(0)$ then $v(s)=u(s)$ for all time $s$, a property known as \textit{balancing} in the literature on learning in feedforward networks \cite{saxe2013exact,atanasov2022alignment,kunin2024get}. Though it would be straightforward and potentially interesting to investigate the effect of unbalanced initialization where $u(s) \neq v(s)$ \cite{kunin2024get}, we will focus on the balanced case as this condition holds approximately at small initialization. 

Assuming balancing, we are left with a two-dimensional system for the learned inverse time constant $c(s)$ corresponding to the outlier eigenvalue $1-c(s)$ and the read-in/read-out scale $u(s) = v(s)$: 
\begin{align}
\begin{split}
\label{E:maindyn1}
\frac{dc(s)}{ds} &= u(s)^2 \left( \frac{u(s)^2}{2 c(s)^2} - \frac{2}{(c(s)+c_\star)^2} \right) \\
\frac{du(s)}{ds} &= - u(s) \left( \frac{u(s)^2}{c(s)} - \frac{2}{c(s)+c_\star} \right)\,,
\end{split}
\end{align}
resulting in an effective filter $f(t) = u^2 e^{-c t}$. We can interpret these two-dimensional dynamics as a gradient flow on an effective loss function 
\begin{align}
L(u, c) = \frac{u^4 }{2 c} - \frac{2 u^2}{c+c_\star} + \frac{1}{2 c_{\star}} 
\end{align}
if we set the learning rate for $u(s)$ to half of that for $c(s)$.

These equations precisely describe the emergence of an outlier eigenvalue, establishing that the final network is a detuned line attractor~\cite{seung1996still}. Through training, $u$ and $1-c$ increase from their $\mathcal{O}(\sigma)$ initial values towards the fixed point of these dynamics at $1-c = 1-c_{\star}$, $u = 1$, corresponding to perfect recovery of the target filter. One can verify by direct computation that this fixed point is stable. Because these dynamics are two-dimensional, we can visualize their flow field directly for a given $c_{\star}$ (Figure \ref{fig:real_target}c). We illustrate the dynamics of~\eqref{E:maindyn1} in Figure~\ref{fig:real_target}b-d, and show an excellent agreement with empirical data. We note that, both before and after learning, activity $\h$ is concentrated in a roughly one-dimensional subspace. This subspace is defined at initialization by the initial read-in vector $\u$, and at the end of learning by the alignment direction $\hat{\u}$ \cite{supplement}.

The two-phase learning dynamics mirror those found for small initialization in linear feedforward networks~\cite{atanasov2022alignment,saxe2013exact}. In the first phase, the read-in and read-out vectors align and grow in magnitude, and in the second phase, the recurrent weights adjust to fine-tune the timescale $c(s)$ toward the target $c_{\star}$. However, the $1/c$--like terms in the effective loss lead to extremely sharp learning curves (Figure~\ref{fig:real_target}d), in contrast to the smooth sigmoids seen in feedforward networks~\cite{saxe2013exact}. As a result, the network can exhibit rapid transitions in performance as $c(s)$ approaches $c_{\star}$. 

Though it is hard in general to analytically pinpoint the time at which this transition to $c(s) = c_{\star}$ occurs, in a simplified model with fixed $u=1$ one can show that as $c_{\ast} \downarrow 0$ the time of convergence scales as $c(0)^{3} \sim (1-\sigma)^{3}$. In particular, if we freeze $u=1$ and train only $c$, the dynamics 
\begin{align}
    \frac{dc}{ds} = \frac{1}{2c^2} - \frac{2}{(c+c_{\star})^2} 
\end{align}
starting from some $c(0) > c_{\star}$ will monotonically decay towards $c_{\star}$ and then stay there. During the decay, one can solve for $s(c)$ and then invert to obtain $c(s)$. The easiest way to summarize the result is in the limit $c_{\star} \downarrow 0$:
\begin{align}
    \lim_{c_{\star} \downarrow 0} c(s) = \begin{cases} 
        \left(c(0)^3 - \frac{9}{2} s \right)^{1/3} & s < \frac{2}{9} c(0)^3,
        \\
        0 & s \geq \frac{2}{9} c(0)^3.
    \end{cases}
\end{align}
Therefore, in this limit the approach to the fixed point becomes nonanalytic. We conjecture that there is a well-defined $c_{\star} \downarrow 0$ limit of the solution to the full dynamics with trainable $u$, but have not endeavored to derive it. This provides a minimal analytical model for past works' observations of long plateaus followed by rapid convergence \cite{haputhanthri2025understanding,li2021curse}.

In simulations, we sometimes observe an overshoot of $c(s)$ beyond the target value $c_\star$. This is consistent with our reduced dynamical model for certain initial conditions. More precisely, $u(s)$ can initially grow fast enough to exceed unity, so that when the RNN time constant $c \approx c_\star$, the derivative $\frac{d c(s)}{ds}$ can be positive at $c = c_\star$. This can be seen by examining the dynamics for $c(s) \approx c_\star$: 
\begin{align}
    \frac{d c(s)}{ds} \approx \frac{1}{2 c_\star^2 }  \ u(s)^2 ( u(s)^2 - 1  ). 
\end{align}
If $u(s) > 1$, this drives overshooting of $c(s)$ beyond $c_{\star}$ until $\frac{d u(s)}{ds}$ becomes sufficiently negative to reduce $u(s)$ closer to unity, which leads $c(s)$ to recover back to $c_{\star}$.

\begin{figure*}[t!]
    \centering
    \includegraphics[width=6.75in]{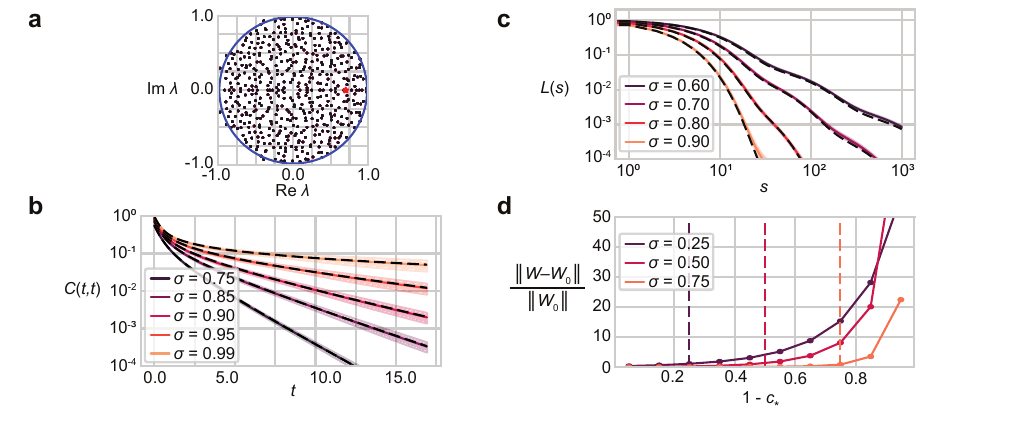}
    \caption{\textbf{Lazily learning to integrate noise.} \textbf{a}. Spectrum of eigenvalues of $\frac{1}{\sqrt{N}} \W(s)$ for a network learning to integrate in the lazy regime, with $\sigma = 0.98$. As in Figure \ref{fig:real_target}, training steps are indicated by color, but no motion of the eigenvalues is visible. The red star indicates the target $1 - c_{\star} = 0.7$. \textbf{b}. The lazy training regime is characterized by the \citet{sompolinsky1988chaos} DMFT autocorrelation function $C(t,t') = \frac{1}{N} \sum_{i=1}^{N} \langle u_{i}(t) u_{i}(t')\rangle$ of the adjoint dynamics at initialization. Here we show the autocorrelation function on the diagonal $t=t'$ estimated from $5$ random initializations of $N=4000$ networks (see SI for smaller values of $N$ which imperfectly concentrate). Theory curves show $C(t,t)=\sigma^2 e^{-2t} I_0(2\sigma t)$. \textbf{c}. Loss dynamics through lazy training in networks of size $N=4000$ with varying initialization variance $\sigma^2$ on a task with $c_\star=0.5$.  Black dashed lines show the dynamics of the loss predicted by the linear dynamics $\partial_s f_s(t) = - \int dt' K(t,t') ( f_s(t') - f_\star(t') )$. \textbf{d}. Relative change in recurrent weights for networks with $N = 250$ as a function of $c_{\star}$. The dashed red vertical line shows the predicted threshold $1-c_{\star} < \sigma$ for lazy learning. Here, we expect the transition to be smoothed both because of finite size effects and because we train a network for a finite amount of time. The effect of early stopping is potentially quite important, as the squared RKHS norm of the learned filter can grow at most linearly with the training time $s$; see the Supplemental Material for details \cite{supplement}. }
    \label{fig:lazy_integrator}
\end{figure*}

\subsection{Dynamics from large initial weights: \texorpdfstring{\\}{} lazy learning}

We next investigate the \emph{lazy learning} regime, where the filter function $f_s(t)$ evolves quickly while the internal recurrent dynamics remain approximately fixed \cite{chizat2019lazy}. This behavior can be induced parametrically either by considering large initialization scales $\sigma$, so that the target eigenvalue $1-c_{\star}$ lies inside the initial reservoir spectrum, or by taking $\gamma \to 0$, which makes the parameter displacement required to substantially change the filter very small. 

Following existing characterizations of lazy learning in generic parametric models \cite{jacot2018neural, chizat2019lazy,alemohammad2020recurrent,yang2021feature,bordelon2022dmft}, we expand the dynamics of $f_s(t)$ by applying the chain rule and using the definition of the updates. This allows us to write 
\begin{align} \label{eqn:kernel_gf}
    \frac{d f_{s}(t)}{ds} = - \int_{0}^{\infty} dt'\, [f_{s}(t')-f_{\star}(t')] K(t,t') ,
\end{align}
where we have defined the neural tangent kernel (NTK)
\begin{align}
    K(t_{1},t_{2}) 
    &= \eta_{\u}\sum_{i=1}^{N} \frac{\partial f(t_{1})}{\partial u_{i}} \frac{\partial f(t_{2})}{\partial u_{i}} + \eta_{\v} \sum_{j=1}^{N} \frac{\partial f(t_{1})}{\partial v_{j}} \frac{\partial f(t_{2})}{\partial v_{j}} \nonumber\\&\quad + \eta_{\W} \sum_{i,j=1}^{N} \frac{\partial f(t_{1})}{\partial W_{ij}} \frac{\partial f(t_{2})}{\partial W_{ij}} . 
\end{align}
In the limit $N \to \infty$, the NTK $K$ is self-averaging with respect to the random initialization of the weights. Moreover, we have that the filter $f_{s=0}(t) \to 0$. If we then take $\gamma \to 0$, $K$ remains constant through training \cite{jacot2018neural, chizat2019lazy,alemohammad2020recurrent,yang2021feature,bordelon2022dmft}. This leaves us with a kernel gradient flow problem. The question is then when the target filter $f_{\star}(t)$ lies in the reproducing kernel Hilbert space (RKHS) generated by $K$. If it does not, then the size of the parameter updates required for $f_{s}(t)$ to match $f_{\star}(t)$ diverges, which means that the network must leave the lazy regime in order to fit the target \cite{jacot2018neural, chizat2019lazy}. We note that this requires $0<\sigma<1$ such that the initial dynamics are stable \cite{supplement,sompolinsky1988chaos,gardiner1985handbook}. 

Our task is then to compute the $N \to \infty$ limit of $K$ at initialization, and determine when the target filter $e^{-c_{\star} t}$ lies in the resulting RKHS. Using the expressions for the derivatives of $f$ in terms of the adjoint dynamics \eqref{eqn:adjoint_dynamics}, we have
\begin{align}
    K(t_{1},t_{2})  &=  \frac{1}{N} \v(t_{1})^{\top} \v(t_{2}) + \frac{1}{N} \u(t_{1})^{\top} \u(t_{2}) \nonumber\\&\quad + \int_{0}^{t_{1}} dt_{3}\, \int_{0}^{t_{2}} dt_{4}\, \left[ \frac{1}{N} \v(t_{3})^{\top} \v(t_{4}) \right] \nonumber\\&\qquad\qquad \times \left[\frac{1}{N} \u(t_{1}-t_{3})^{\top} \u(t_{2}-t_{4}) \right],
\end{align}
where we drop the implied $s=0$ training time subscript on the adjoint variables. Thus, we must compute the limiting autocorrelations of the adjoint dynamics $\u(t)$ and $\v(t)$ at initialization. To do so, we first note that, as $\v$ and $\u$ are identically distributed at initialization, their adjoint dynamics are identically distributed, thanks also to the fact that $\W$ and $\W^{\top}$ are identically distributed. As a simple check on the answer, we should have that $\lim_{N \to \infty} K(0,0) = 2 \sigma^{2}$. 

Therefore, it is sufficient to analyze the read-in adjoint dynamics 
\begin{align}
    \frac{d}{dt} \u(t) = - \u(t) + \frac{1}{\sqrt{N}} \W \u(t), \quad \u(t=0) = \u, 
\end{align}
where we recall that the elements of the matrix $\W$ and the initial vector $\u$ are all i.i.d. $\mathcal{N}(0,\sigma^2)$. This is itself a random linear RNN, whose dynamics can be analyzed using a standard dynamical mean-field theory (DMFT) approach, following \citet{sompolinsky1988chaos}. As $N \to \infty$, the elements of the adjoint vector statistically decouple, and the interactions $\frac{1}{\sqrt{N}} \W \u(t) $ generate an effective Gaussian noise in the dynamics for $\u(t)$: 
\begin{align}
    (1 + \partial_t ) u_i(t) = \xi_i(t).
\end{align}
Here, $\xi_{i}(t)$ is a zero-mean Gaussian process with covariance 
\begin{align}
    \langle \xi_i(t) \xi_j(t') \rangle = \delta_{ij} \sigma^2  C(t,t') 
\end{align}
where $C(t,t')$ is the self-consistently determined autocorrelation function
\begin{align}
    C(t,t') = \frac{1}{N} \sum_{i=1}^{N} \langle u_{i}(t) u_{i}(t') \rangle.
\end{align}
In the $N \to \infty$ limit, the autocorrelation $C(t,t')$ converges to a deterministic (initialization-independent) function. This function satisfies the differential equation
\begin{align}
    (1 + \partial_t) (1+\partial_{t'}) C(t,t') = \sigma^2 C(t,t') 
\end{align}
for $t,t' \geq 0$, with boundary conditions $C(0,0) = \sigma^{2}$ and $C(t,0) = C(0,t) = \sigma^{2} e^{-t}$ following from the definition of the dynamics \cite{sompolinsky1988chaos}. We show in Appendix~\ref{app:dmft_solution} that this equation can be solved explicitly in terms of the modified Bessel function of the first kind $I_0(\cdot)$:  
\begin{align}
    C(t,t') = \sigma^2 e^{-t - t'} I_{0}(2 \sqrt{ \sigma^2 t t'}) .
\end{align}
This analytical solution yields an excellent match to simulations on a per-realization basis (Figure~\ref{fig:lazy_integrator}b). 

Therefore, we find that 
\begin{align}
    \frac{1}{N} \v(t_{1})^{\top} \v(t_{2}),\, \frac{1}{N} \u(t_{1})^{\top} \u(t_{2}) \to C(t_{1},t_{2})
\end{align}
as $N \to \infty$. In turn, this allows us to write down an explicit solution for the NTK (see Appendix~\ref{app:dmft_solution}):
\begin{align}
\begin{split}
    K(t,t') = \sigma^{2} e^{-t-t'} ( 2 I_{0}(2 \tau) + \tau I_{1}(2\tau) ) , 
\end{split}
\end{align}
where we put $\tau = \sqrt{\sigma^{2} tt'}$ for brevity. We can then numerically solve the dynamics \eqref{eqn:kernel_gf} to predict the loss function, and find good matches to finite-size simulations (Figure~\ref{fig:lazy_integrator}c). 

With this solution in hand, we can determine when the target filter $f(t) = e^{-c_{\star} t}$ is lazily learnable in the $N \to \infty$ limit. In Appendix~\ref{app:rkhs}, we show that there is a sharp and intuitive threshold condition: the target is in-RKHS if and only if
\begin{align}
    1 - \sigma < c_{\star} < 1 + \sigma, 
\end{align}
that is, if the required inverse timescale lies within the limiting spectrum of $\mathbf{I} - \frac{1}{\sqrt{N}} \W$, which is a disk of radius $\sigma$ centered at 1. Finite size experiments at non-vanishing $\gamma$ show an increase in the parameter displacement as $1-c_{\star}$ crosses $\sigma$, consistent with this limiting result (Figure~\ref{fig:lazy_integrator}d). In other words, if the target timescale lies far outside the initial bulk, the network must adapt $\W$ in order to fit it. This is consistent with recent proposals for when rich learning in feedforward networks can accelerate training \cite{bordelon2024improve}. 

\section{Learning an oscillator} 

Next we study learning of a richer memory task: integration with a damped sinusoidal filter
\begin{align}
    f_{\star}(t) = e^{-c_{\star} t} \cos(\omega_{\star} t). 
\end{align}
This target filter represents an oscillatory memory trace with frequency $\omega_{\star}$ and damping rate $c_{\star}$, requiring the network to develop oscillatory dynamics as it learns. As was the case for the integrator, the nature of the learning dynamics strongly depends on the initialization scale $\sigma$. If $\sigma \simeq 1$, the network can lazily learn to match the oscillatory filter much as it lazily learned to integrate (Figure \ref{fig:lazy_oscillator_supp}). As in the real decay case, we can prove a sharp threshold for lazy learnability depending on the spectral radius of the initial weights: the oscillating decay can be learned in the lazy regime if and only if $\sqrt{(1-c_{\star})^{2} + \omega_{\star}^{2}} < \sigma$ (Appendix~\ref{app:osc_in_rkhs}).

\begin{figure*}[t!]
    \centering
    \includegraphics[width=7in]{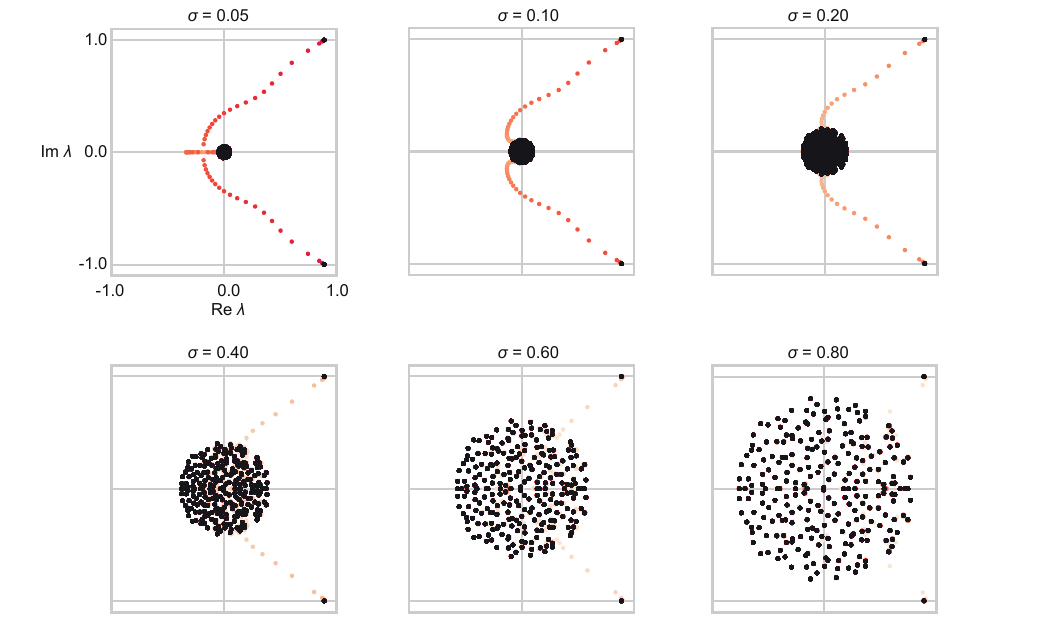}
    \caption{\textbf{Learning an oscillator.} We simulate networks with $N=250$ learning to emulate an oscillatory filter $f_{\star}(t) = e^{-c_{\star} t} \cos(\omega_{\star} t)$ with parameters $(c_{\star},\omega_{\star}) = (0.1,1.0)$, starting from initializations of different scale $\sigma$. Training time is indicated by color, with darker colors indicating later times. In all cases, $\sigma < \sqrt{(1-c_{\star})^2 + \omega_{\star}^2}$, so the target eigenvalues do not lie within the initial spectral support, and $f_{\star}$ is not lazily learnable. We see that the escape path for the pair of outlier eigenvalues is highly dependent on the initialization scale $\sigma$. In the very rich regime (here in the case $\sigma = 0.05$), we see a particularly striking phenomenon: First one outlier eigenvalue emerges along the negative real axis, then a second. The two eigenvalues then collide and move off the real axis, evolving from that point on as a complex-conjugate pair. Therefore, unlike the simple decaying target filter, capturing the geometry of the initial gradient flow dynamics at non-negligible $\sigma$ would require computing complicated interactions from the bulk eigenvalues, which we leave for future work.}
    \label{fig:rich_oscillator_supp}
\end{figure*}

If the weights are small ($\sigma \ll \sqrt{ (1-c_{\star})^{2} + \omega_{\star}^{2}}$), the learning dynamics consists of an alignment phase, after which subsequent learning is described by a reduced-dimensional effective dynamics. In Figure \ref{fig:rich_oscillator_supp} we illustrate the emergence of two outlier eigenvalues that eventually converge to the points $\lambda_\star = 1-c_\star \pm i\omega_\star$ in the complex plane. We note that the escape geometry of these outliers depends critically on $\sigma$, unlike the real target case. In particular, if $\sigma \downarrow 0$, then we in fact observe the sequential emergence of two eigenvalues along the negative real axis, which then collide and move off-axis. These dynamics are substantially more challenging to analyze than the alignment dynamics in the real case. Therefore, we focus our attention on the reduced dynamics in the post-alignment phase. 

After alignment, the modes responsible for the oscillatory behavior at small $\sigma$ span a two-dimensional real subspace of the full $N$-dimensional dynamical space. We now show that we can reduce the full $N(N+2)$-dimensional parameter space gradient flow dynamics to an 8-dimensional space of effective parameters defined by projecting into the 2-dimensional dynamical subspace. Let $\mathbf P \in \mathbb{R}^{N \times N}$ be an orthogonal projection to the subspace spanned by the oscillatory modes, satisfying $\mathbf P^2 = \mathbf P$. The confinement of the dynamics to the subspace is equivalent to the statement that $\u, \v, \W$ are (nearly) unchanged under projection:
\begin{align}
     \mathbf P \u_{s} \approx \u_{s} , \  \mathbf P \v_{s} \approx \v_{s} , \ \mathbf P \W_{s} \mathbf P \approx \W_{s}  .
\end{align}
Under this alignment condition, we can verify that the gradients are also confined to the same subspace:
\begin{align}
    &\frac{d \W_s}{ds} \propto \mathbf P \frac{d\W_s}{ds}  \mathbf P,\ \frac{d \u_s}{ds} \propto \mathbf P \frac{d \u_s}{ds}, \ \frac{d \v_s}{ds} \propto \mathbf P  \frac{d \v_s}{ds}. 
\end{align}

Therefore, restricting our attention to the projected components, we have an 8-dimensional set of effective parameters:
\begin{align}
    \begin{pmatrix}
        u_1(s)
        \\
        u_2(s)
    \end{pmatrix} , \ \begin{pmatrix}
        v_1(s)
        \\
        v_2(s)
    \end{pmatrix} , \ \begin{pmatrix}
        W_{11}(s) & W_{12}(s)
        \\
        W_{21}(s) &  W_{22}(s)
    \end{pmatrix}. 
\end{align}
The parameters $W_{11},W_{12},W_{21},W_{22}$ represent the effective recurrent weights, while $u_1,u_2$ and $v_1,v_2$ represent the read-in and read-out weights. These parameters define a set of reduced dynamics described by a two-dimensional \textit{effective} linear RNN in the 2D subspace: 
\begin{align}
\label{E:reduced1}
    \dot{\mathbf{h}}_{\text{eff}}(t) &= - \mathbf{h}_{\text{eff}}(t) + \begin{pmatrix} W_{11} & W_{12} \\ W_{21} & W_{22} \end{pmatrix} \mathbf{h}_{\text{eff}}(t) + \begin{pmatrix} u_1 \\ u_2 \end{pmatrix} x(t) , 
\end{align}
which leads to an effective filter
\begin{equation}
    f_{\text{eff}}(t) = \frac{1}{\gamma} \begin{pmatrix} v_1 & v_2 \end{pmatrix}^{\!} \exp\left\{\begin{pmatrix} W_{11}-1 & W_{12} \\ W_{21} & W_{22} - 1 \end{pmatrix} t\right\} \begin{pmatrix} u_1 \\ u_2 \end{pmatrix}\,.
\end{equation}

To circumvent the challenge of predicting escape dynamics, we initialize the network with a warm start that initializes two outlier eigenvalues outside of the bulk. We accomplish this by setting
\begin{align}
    \frac{1}{\sqrt N} \W = \frac{1}{\sqrt N} \W_0 + c_+ \z \z^\dagger + c_{-} \z_\ast \z^\top. 
\end{align}
Here, $\z \in \mathbb{C}^{N}$ is a complex vector, with element-wise conjugate $(z_{\ast})_{j} = z_{j}^{\ast}$ and Hermitian conjugate $\z^{\dagger} = \z_{\ast}^{\top}$. $\z$ and $\z_{\ast}$ are then approximate eigenvectors of $\W$ with eigenvalues $\{ c_{\pm} \}$, which we set with norm $|c_{\pm}|^2 > \sigma^2$ so that they lie outside of the bulk. The vector $\z$ is initialized with orthonormality constraints $\z^\top \z = 0$ and $\z^\dagger \z = 1$. We also initialize the read-in and read-out with an aligned warm start as $\v = \u = \frac{\sigma \sqrt N}{\sqrt 2} (\z + \z_\ast) \in \mathbb{R}^N$. Then, the two-dimensional subspace of interest is the span of $\{ \z \pm \z_\star \}$

\begin{figure}[t!]
    \centering
    \setlength{\belowcaptionskip}{-12pt}
    \includegraphics[width=3.38in]{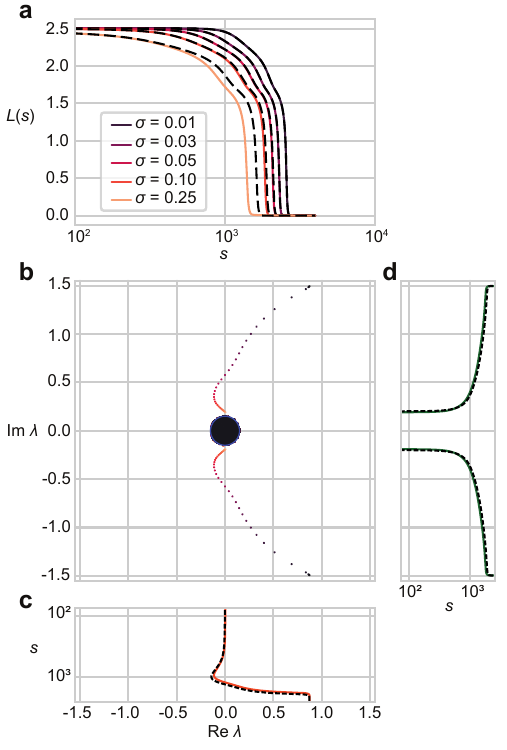}
    \caption{\textbf{Learning to mimic oscillatory filters in the rich regime.} The dynamics of an RNN with $N = 250$ hidden units learning an oscillatory filter with parameters $(c_\star, \omega_\star ) = (0.1,1.0)$ can be approximated with the dynamics of a $2$ neuron RNN. (a) The losses as a function of $s$ and initialization scale $\sigma$. Larger $\sigma$ leads to more bulk/outlier interaction causing deviations. (b)-(d) Outlier dynamics under a warm start. in (b) we show the outlier dynamics in one simulation, with time indicated by shading. In (c) and (d), we show the trajectories of the real and imaginary parts of the two outliers, respectively. Solid lines show experiment, and dashed lines show the predictions generated from the two-neuron effective RNN. }
    \label{fig:complex_target}
\end{figure}

It is then straightforward to simulate the learning dynamics of our effective two-neuron model and compare it to an $N$-neuron network (Figure~\ref{fig:complex_target}). The dynamics of the $N$ neuron system reveal that the bulk eigenvalues are effectively unchanged, but the outliers follow a complicated dynamics as they approach their final values $\lambda_\star= 1-c_\star \pm i \omega_\star$. These outlier eigenvalues correspond to the emergent oscillatory mode that the network learns to match the target filter. The complicated dynamics of these eigenvalues is quantitatively reproduced by our effective model. The close agreement between the effective model and the full network underscores the power of reduced-dimensional analyses when learning tasks with effectively low dimensional structure. When the outliers are sufficiently far from the bulk, the dynamics are well captured by the two-dimensional RNN.

\section{Towards multi-timescale targets}

\begin{figure}
    \centering
    \includegraphics[width=3.38in]{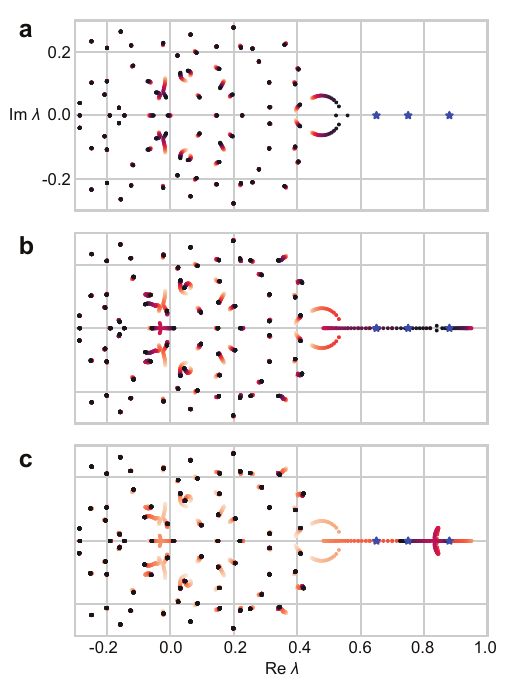}
    \caption{\textbf{Evolution of outlier eigenvalues on a task with three real target timescales.} Two outliers emerge, collide (a), separate (b), and collide again (c) before converging to their final values on the real axis. The dynamics are challenging to understand through snapshots alone. We therefore make videos of these collisions available on GitHub at \url{https://github.com/Pehlevan-Group/rnn-learning-dynamics-theory}.}
    \label{fig:multiple_real_times}
\end{figure}

Going beyond learning filters $f_\star(t)$ with a single target time-constant, we next consider more general filters which are linear combinations of exponentials with timescales $\{ c_n \}$ with coefficients $\{ a_n \}$
\begin{align}
    f_\star(t) = \sum_{n} a_n e^{- c_n t} . 
\end{align}
In this task, the internal recurrent weights must encode multiple timescales and we expect from the preceding analysis that the rich regime would result in multiple outliers emerging. However, simulations suggest that exact characterization of the outlier dynamics is challenging. In Figure \ref{fig:multiple_real_times} we illustrate a filter with three real target timescales $c_n \in \mathbb{R}$. Gradient flow illustrates very nontrivial dynamics of the outliers. At first, two outliers emerge in the complex plane (Figure \ref{fig:multiple_real_times}a), they collide on the real axis, separate (Figure \ref{fig:multiple_real_times}b), collide again into the complex plane, then return to the real axis (Figure \ref{fig:multiple_real_times}c). These nontrivial collision and separation events motivate further analysis but are beyond the scope of our present analytical tools. 

\section{Discussion}

We have analyzed the dynamics of learning to integrate via gradient descent in a streamlined RNN model. By focusing on linear RNNs and specific target filters, we derived analytical expressions that precisely describe how network parameters evolve during training. These solutions reveal two distinct learning regimes: a rich regime where the read-in, read-out, and recurrent weights first align and then grow in scale to match the desired temporal dynamics, and a lazy regime in which timescales from the initial reservoir are mixed to mimic the target. The transition between these regimes is controlled by the scale of the initial weights, and by the target timescale. Our analytical results illuminate the conditions governing RNN operation in rich versus lazy learning regimes, particularly how weight initialization scale and target memory timescale determines these dynamics. This understanding is crucial for designing networks that effectively learn long-timescale dependencies.

Our work is, however, just a first step. We have focused on learning very low-dimensional integration tasks for uncorrelated inputs by minimizing the population loss; accounting for the effects of finite training data and temporal correlations in input signals will be required to understand how RNNs learn to solve richer dynamical problems. Moreover, we have restricted our attention to linear RNNs. This focus is motivated by experience from the study of learning dynamics in feedforward networks, where linear networks have been shown to capture salient phenomenology in the nonlinear dynamics of optimization \cite{saxe2013exact,fukumizu1998effect,kunin2024get,atanasov2022alignment}. Detailed experimental tests of the degree to which phenomena we observe in the linear RNN setting transfer to nonlinear RNNs will be an important objective for future work. At the same time, linear RNNs are not just a toy model from the perspective of modern machine learning, as state-space sequence models are based directly on linear RNNs \cite{orvieto2023resurrecting,zucchet2024recurrent,gu2024mamba,gu2020hippo}.

Even so, our results show that training a linear RNN to integrate white noise or learn a damped oscillatory filter leads to the spontaneous emergence of topological structures such as detuned line attractors or oscillatory modes. From a dynamical systems perspective, this suggests that gradient-based learning sculpts the topology of dynamics on the network’s phase space, a principle that has been explored in nonlinear settings~\cite{sussillo2013box, maheswaranathan2019reverse,maheswaranathan2019individuality,cotler2023analyzing,driscoll2024flexible, cotler2024computational}. Our findings thus reinforce the idea that trained recurrent networks naturally develop low-dimensional attractor structures for memory and temporal processing, as observed in both artificial and biological systems~\cite{sussillo2013box,vayas2020computation,khona2022attractor,maheswaranathan2019reverse,maheswaranathan2019individuality}. 

Our work builds upon a line of research by Ostojic, Barak, and colleagues on RNNs with low-rank structure in their recurrent weights \cite{mastrogiuseppe2018linking,dubreuil2022population,susman2021quality,schuessler2020interplay,schuessler2024oblique}. In \citet{schuessler2024oblique}, those authors identified ``aligned'' and ``oblique'' regimes of operation in RNNs, defined by the alignment of the recurrent weight matrix with the read-out direction. They demonstrated that lazy learning leads to ``oblique'' dynamics, while rich learning drives alignment. Those results build on their earlier work \cite{schuessler2020interplay}, where they analyzed gradient flow learning in linear RNNs for tasks that depend only on the long-time behavior, \textit{i.e.}, on the fixed-point output $y(t \to \infty) = \frac{1}{N} \v^{\top} (\mathbf{I} - \frac{1}{\sqrt{N}} \W)^{-1} \u\, x(t\to\infty)$. Our work complements this prior art by providing a setting where one can in fact derive a complete prediction for the learning trajectory for dynamical tasks, at the expense of the restriction to low-dimensional tasks and linear dynamics. Our results illuminate in detail how the aligned dynamics observed by \citet{schuessler2024oblique} emerge through learning. 

We emphasize that, though we focus on low-dimensional integration tasks, the structure of the tasks themselves does not enforce the emergence of low-dimensional recurrent dynamics. Rather, whether or not low-dimensional dynamics emerge depends on the interaction of the task with the inductive bias of gradient flow starting from Gaussian initialization. The reservoir dynamics of the lazy regime show that, for a fixed target timescale, one can alter the dimensionality of the final learned dynamics by changing the initialization scale. Moreover, from the perspective of filter approximation alone, there exist high-dimensional solutions that achieve arbitrarily low loss. Concretely, we show in Appendix \ref{app:chain} how one can construct feedforward chains---\textit{i.e.}, networks with $W_{ij} \propto \delta_{i,j+1}$---that achieve exponentially-decaying loss as a function of network size. Mapping out how learning rules interact with structure in initial weights to determine what integration mechanisms are learned will be an important goal for future work \cite{fiete2010stdp,liu2024how,bahle2024neural,qian2026discovering}.

From a theoretical perspective, our two-phase approach to analyzing learning dynamics in the rich regime is not without limitations. Conceptually, perhaps the most important among these limitations is the fact that the complexity of the reduced dynamics following alignment increases substantially for tasks beyond leaky integration. We saw this already in the case of learning a damped sinusoid, where the natural reduction is to a two-dimensional linear RNN. Compared to the two-dimensional effective dynamics for leaky integration, this forces us to analyze a six-dimensional dynamical system. Moreover, as we saw both for the damped sinusoid and for the multi-timescale filters considered in Figure~\ref{fig:multiple_real_times}, the learning dynamics for more complicated filters include strong interactions such as eigenvalue collisions. These could in principle be studied analytically, and doing so will be an interesting goal for future work. 

Another limitation of our analysis as it stands is that we ignore interactions between the low-dimensional set of slow modes and the high-dimensional bulk of eigenvalues. Accounting for these interactions could allow us to track the dynamics in the damped sinusoid task without assuming alignment has already occurred, as we conjecture that the scale-dependent escape geometry we observed in Figure~\ref{fig:rich_oscillator_supp} results in part from these interactions. The natural approach to incorporating the effect of the bulk on the dynamics of the outlier eigenvalues is through dynamical mean-field theory, as has been recently done for feedforward networks by \citet{lauditi2026spectral}. This would result in an asymptotically exact description, though one given in terms of equations of not inconsiderable complexity \cite{lauditi2026spectral,clark2026structure}. 

Looking forward, developing a detailed theoretical understanding of how RNNs learn to solve simple tasks is an important prerequisite to identifying the mechanisms underlying computation through neural dynamics \cite{vayas2020computation}. In confronting theories for integration mechanisms with data \cite{daie2023feedforward,khona2022attractor,qian2024partial,oshea2022direct} and dissecting natural constraints on learning dynamics \cite{sadtler2014constraints,oby2025dynamical}, it is important to keep in mind inductive biases of simple RNNs as a baseline.

\section{Acknowledgements} 

We thank William Qian for helpful comments on a previous version of this manuscript. J.A.Z.-V. thanks M. Fee and I. Fiete for stimulating discussions regarding structured initializations. B.B. is supported by a Google PhD Fellowship. C.P. is supported by an NSF CAREER Award (IIS-2239780), DARPA grants DIAL-FP-038 and AIQ-HR00112520041, the Simons Collaboration on the Physics of Learning and Neural Computation, and the William F. Milton Fund from Harvard University. This work has been made possible in part by a gift from the Chan Zuckerberg Initiative Foundation to establish the Kempner Institute for the Study of Natural and Artificial Intelligence. J.A.Z.-V. is supported by the Office of the Director of the National Institutes of Health under Award Number DP5OD037354. The content is solely the responsibility of the authors and does not necessarily represent the official views of the National Institutes of Health. JAZV is further supported by a Junior Fellowship from the Harvard Society of Fellows.

\section{Author contributions} 

Conceptualization, B.B., J.C., and J.A.Z.-V., with input from C.P.; Methodology, B.B., J.C., and J.A.Z.-V., with input from C.P.; Investigation, B.B., J.C., and J.A.Z.-V.; Writing – Original Draft, B.B., J.C., and J.A.Z.-V.; Writing – Review \& Editing, B.B., J.C., C.P., and J.A.Z.-V.; Funding Acquisition, C.P. and J.A.Z.-V.

\section{Code availability} 

All code was implemented in Python using JAX \cite{jax2018github}, and is available on GitHub at: \url{https://github.com/Pehlevan-Group/rnn-learning-dynamics-theory}.


\appendix 

\section{Solution to the DMFT equations}\label{app:dmft_solution}

Here, we show that the DMFT equation 
\begin{align}
    (1 + \partial_t) (1+\partial_{t'}) C(t,t') = \sigma^2 C(t,t') 
\end{align}
for $t,t' \geq 0$, with initial condition $C(0,0) = \sigma^{2}$ and boundary condition $C(t,0) = C(0,t) = \sigma^{2} e^{-t}$, can be solved explicitly in terms of the modified Bessel function of the first kind $I_0(\cdot)$, yielding 
\begin{align}
    C(t,t') = \sigma^2 e^{-t - t'} I_{0}(2 \sqrt{ \sigma^2 t t'}) .
\end{align}
This solution is easy to verify using identities for $I_0$. As $I_{0}(0) = 1$, the initial condition is clearly satisfied. Now, for any $t,t'>0$, we observe that
\begin{align}
    (1+\partial_{t'}) C(t,t') 
    &= \sigma^{2} e^{-t-t'} \partial_{t'} I_{0}(2 \sqrt{\sigma^2 tt'})
    \\
    &= \sigma^{2} e^{-t-t'}I_{1}(2 \sqrt{\sigma^2 tt'}) \sqrt{\frac{t \sigma^2}{t'}}, 
\end{align}
so 
\begin{align}
\begin{split}
    &(1+\partial_{t}) (1+\partial_{t'}) C(t,t') 
    \\
    &\quad =  \sigma^{2} e^{-t-t'} \frac{1}{2} \bigg[ I_{0}(2 \sqrt{\sigma^2 tt'}) + I_{2}(2 \sqrt{\sigma^2 tt'}) \\& \qquad\qquad\qquad\quad +  I_{1}(2 \sqrt{\sigma^2 tt'}) \sqrt{\frac{1}{\sigma^2 t t'}}  \bigg] \sigma^2 .
\end{split}
\end{align}
But, using the identity
\begin{align}
    \frac{2\nu}{x} I_{\nu}(x) = I_{\nu-1}(x) - I_{\nu+1}(x),
\end{align}
we can simplify this to
\begin{align}
    (1+\partial_{t}) (1+\partial_{t'}) C(t,t') &= \sigma^{4} e^{-t-t'} I_{0}(2\sqrt{\sigma^2 t t'}) \\&= \sigma^{2} C(t,t'),
\end{align}
which proves the claim. 

We now prove that this expression for $C$ results in the claimed expression for the NTK $K$. Recall that 
\begin{align}
    K(t_{1},t_{2}) &= 2 C(t_{1},t_{2}) + C^{\ast 2}(t_{1},t_{2}),
\end{align}
where we have defined the convolution
\begin{align}
    C^{\ast 2}(t_{1},t_{2}) = \int_{0}^{t_{1}} dt_{3}\, \int_{0}^{t_{2}} dt_{4}\, C(t_{3},t_{4})  C(t_{1}-t_{3}, t_{2}-t_{4}).
\end{align}
We claim that
\begin{align}
    C^{\ast 2}(t_{1},t_{2}) = \sigma^{2} e^{-t_{1}-t_{2}} \sqrt{\sigma^2 t_{1} t_{2}} I_{1}(2 \sqrt{\sigma^{2} t_{1} t_{2}}) . 
\end{align}
To show this, we start with the series expansion of the Bessel function expression for $C(t,t')$: 
\begin{align}
    C(t,t') = \sigma^2 e^{-t - t'} \sum_{n=0}^{\infty} \frac{(\sigma^{2} tt')^{n}}{(n!)^2} . 
\end{align}
As 
\begin{align}
    \int_{0}^{t_{1}} t_{3}^{n} (t_{1}-t_{3})^{m}\,dt_{3} &= \frac{t_{1}^{n+m+1} m! n!}{(m+n+1)!}, 
\end{align}
integrating term-by-term allows us to write the convolution as
\begin{align}
\begin{split}
    C^{\ast 2}(t_{1},t_{2})
    &= \sigma^{2} e^{-t_{1}-t_{2}} \sum_{n,m=0}^{\infty} \frac{(\sigma^{2} t_{1}t_{2})^{m+n+1}}{[(m+n+1)!]^2} 
    \\
    &= \sigma^{2} e^{-t_{1}-t_{2}} \sum_{k=1}^\infty  k \  \frac{(\sigma^2 t_1 t_2)^{k}}{(k!)^2}  
    \\
    &= \sigma^{2} e^{-t_{1}-t_{2}} \sqrt{\sigma^{2} t_{1}t_{2}} I_{1}(2\sqrt{\sigma^2 t_{1}t_{2}}),
\end{split}
\end{align}
in terms of another modified Bessel function \cite{dlmf}. This is the desired result.

\section{Thresholds for lazy learnability}\label{app:rkhs}

In this Appendix, we prove the claimed thresholds for lazy learnability. These thresholds depend on the notion of a reproducing kernel Hilbert space (RKHS). We first recall the definition of the RKHS norm in terms of a general feature space \cite{berlinet2004rkhs}, letting the NTK $K$ stand in place of a general positive-definite kernel function. Given a sequence of functions $\alpha_{n}(t)$, usually referred to as \emph{features}, such that
\begin{align}
   K(t,t') = \sum_{n=0}^{\infty} \alpha_{n}(t) \alpha_{n}(t'),
\end{align}
the RKHS norm of some function $g$ is precisely the minimum-norm sequence of coefficients that allows us to express the function in terms of these features: 
\begin{align}
    \Vert g \Vert_{K}^{2} = \inf_{\{c_{n}\}}\left\{ \sum_{n=0}^{\infty} |c_{n}|^{2} \,\middle\vert\, \sum_{n=0}^{\infty} c_{n} \alpha_{n}(t) = g(t) \right\} .
\end{align}
If no such decomposition is possible, then the norm is infinite, and the function $g$ lies outside the RKHS generated by the kernel $K$. Note that this definition does not require the functions $\alpha_{n}(t)$ to be orthogonal. 

For kernel gradient flows of the form \eqref{eqn:kernel_gf}, the RKHS norm of the target function $f_{\star}(t)$ determines whether the norm of the parameter displacement---in the sense of the feature coefficients of $f(t)$---must diverge in order to fit $f_{\star}(t)$ \cite{supplement}. This is why the RKHS norm defines the threshold for lazy learnability. 

In the Supplemental Material \cite{supplement}, we approach the same problem from a different perspective: instead of considering the feature-space expansion of the kernel, we consider its Mercer eigendecomposition \cite{berlinet2004rkhs}. This is more analytically intensive, but provides more detailed information as the training dynamics of the filter admit a simple expansion in terms of the orthonormal set of Mercer eigenvectors. For $C$, we can show that the eigenvectors are related to Laguerre polynomials thanks to the Hardy-Hille formula for Bessel functions \cite{alsalam1964laguerre}, and in so doing recover the same threshold for learnability of exponential decays. However, the NTK $K$ does not admit a simple eigendecomposition. Rather, the best we can do is to show that it is tridiagonal in the basis of eigenvectors of $C$. Therefore, here we restrict our attention to the feature-space view. 

\subsection{Norms of exponential decays in the RKHS generated by \texorpdfstring{$C$}{C}}

We first consider the RKHS generated by the autocorrelation function $C$, as a prelude to that generated by $K$. This space has a natural interpretation: it is the space of functions learnable through reservoir computing---those that could be learned if one froze $\u$ and $\W$ and trained only the read-out vector $\v$ \cite{qian2025fundamental,alemohammad2020recurrent,bordelon2022population}. 

Using the series expansion of the Bessel function, we have
\begin{align}
    C(t,t') = \sigma^2 e^{-t - t'} \sum_{n=0}^{\infty} \frac{(\sigma^{2} tt')^{n}}{(n!)^2} , 
\end{align}
which gives us features
\begin{align}
    \alpha_{n}(t) = \frac{1}{n!} e^{-t} \sigma^{n+1} t^{n} . 
\end{align}
But, using the series expansion of the exponential function, we have
\begin{align}
    e^{-c_{\star} t} 
    &= e^{-t} e^{(1-c_{\star}) t} = e^{-t} \sum_{n=0}^{\infty} \frac{(1-c_{\star})^{n} t^{n}}{n!},
\end{align}
which leads to the decomposition
\begin{align}
    e^{-c_{\star} t} = \sum_{n=0}^{\infty} \frac{(1-c_{\star})^{n}}{\sigma^{n+1}} \alpha_{n}(t) ,
\end{align}
which is unique because the exponential function is entire. Therefore, the exponential decay is in-RKHS if
\begin{align}
    \frac{1}{\sigma^{2}} \sum_{n=0}^{\infty} \left(\frac{(1-c_{\ast})^2}{\sigma^2} \right)^{n} < \infty, 
\end{align}
which holds precisely if $1 - \sigma < c_{\star} < 1 + \sigma$. Moreover, we can compute that
\begin{align}
    \Vert e^{-c_{\star} t} \Vert_{C}^{2} = \frac{1}{\sigma^{2} - (1-c_{\star})^2} . 
\end{align}

\subsection{Norms of exponential decays in the RKHS generated by the NTK}

Now we turn our attention to the NTK. Expanding the two Bessel functions that appear and combining terms, 
\begin{align}
    K(t,t') = \sigma^{2} e^{-t-t'}  \sum_{n=0}^{\infty} (2+n) \frac{(\sigma^{2} tt')^{n}}{(n!)^2}. 
\end{align}
Therefore, we can represent $K$ in terms of the features
\begin{align}
    \tilde{\alpha}_{n}(t) = \frac{1}{n!} e^{-t} \sqrt{2 + n} \sigma^{n+1} t^{n}. 
\end{align}
This gives us the decomposition
\begin{align}
    e^{-c_{\star} t}
    &= \frac{1}{\sigma} \sum_{n=0}^{\infty} \frac{(1-c_{\star})^{n}}{\sqrt{2+n} \sigma^{n}} \tilde{\alpha}_{n}(t),
\end{align}
which means that the exponential decay is in-RKHS if
\begin{align}
    \frac{1}{\sigma^{2}} \sum_{n=0}^{\infty} \frac{(1-c_{\star})^{2n}}{(2+n) \sigma^{2n}} < \infty. 
\end{align}
By direct comparison, the convergence of this series is equivalent to convergence of the series
\begin{align}
    \sum_{n=1}^{\infty} \frac{1}{n} \left(\frac{(1-c_{\ast})^2}{\sigma^2} \right)^{n} ,
\end{align}
which converges whenever $1 - \sigma < c_{\star} < 1 + \sigma$, and diverges otherwise. This is the same condition we found for the RKHS generated by $C$. The full sum, however, does not have an elementary form (it can be expressed in terms of hypergeometric functions, but we will not do so as it is not illuminating). 

\subsection{Which oscillating exponential decays are in-RKHS?}\label{app:osc_in_rkhs}

We now extend the results of the preceding sub-sections to decaying sinusoids
\begin{align}
    f_{\star}(t) = e^{-c_{\star} t} \cos(\omega_{\star} t) . 
\end{align}
As in the pure-decay case, we expand the target filter as a power series: 
\begin{align}
    f_{\star}(t) 
    &= e^{-t} \frac{1}{2} (e^{(1-c_{\star} + i \omega_{\star}) t} + e^{(1 - c_{\star} - i \omega_{\star}) t})
    \\
    &= e^{-t} \sum_{n=0}^{\infty} \frac{t^{n}}{n!} \frac{(1-c_{\star} + i \omega_{\star})^{n} + (1-c_{\star} - i \omega_{\star})^{n}}{2}. 
\end{align}

For $C$, this allows us to identify expansion coefficients
\begin{align}
    f_{\star}(t) = \frac{1}{\sigma} \sum_{n=0}^{\infty} \Re(z^{n}) \alpha_{n}(t),
\end{align}
where we set
\begin{align}
    z = \frac{1-c_{\star} + i \omega_{\star}}{\sigma}. 
\end{align}
This means that the function is in the RKHS generated by $C$ if
\begin{equation}
    \infty > \frac{1}{\sigma^2} \sum_{n=0}^{\infty} \left| \Re(z^{n}) \right|^{2} 
    = \frac{1}{\sigma^2} \sum_{n=0}^{\infty} |z|^{2n} \cos(n \arg z)^2.
\end{equation}
By direct comparison, this series converges if and only if $|z| < 1$, which translates to the condition $\sqrt{(1-c_{\star})^2+\omega_{\star}^2} < \sigma$. In terms of $z$, we can write the RKHS norm compactly as 
\begin{align}
    \Vert f_{\star} \Vert_{C}^{2} = \frac{1}{2 \sigma^{2}} \left[\frac{1}{1-|z|^2} + \Re\left( \frac{1}{1-z^2} \right) \right] 
\end{align}
when $|z|<1$. 

Considering the NTK $K$, we have a similar expansion 
\begin{align}
    f_{\star}(t) &= \frac{1}{\sigma} \sum_{n=0}^{\infty} \frac{\Re(z^{n})}{\sqrt{2+n} \sigma^{n}} \tilde{\alpha}_{n}(t),
\end{align}
which leads to the condition
\begin{align}
    \infty &> \frac{1}{\sigma^2} \sum_{n=0}^{\infty} \frac{|z|^{2n}}{2+n} \cos(n \theta)^2. 
\end{align}
This again converges if and only if $|z|<1$. As in the pure-decay case, the RKHS norm $\Vert f_{\star} \Vert_{K}^{2}$ does not have an illuminating closed form. 

\section{High-rank solutions with low loss}\label{app:chain}

We found that, in the rich regime, gradient descent selects a low-rank solution. One might ask whether this effect is purely driven by the tasks we selected, \textit{i.e.}, if these low-rank solutions are the only ways to achieve low loss. In this Appendix, we emphasize that this is not the case: there exist high-rank solutions that achieve exponentially-decaying loss in $N$.

Consider a pure feedforward chain, in which all input is passed into the first neuron, 
\begin{align}
    u_{j} = \delta_{j,1},
\end{align}
and then propagated along the chain via
\begin{align}
    W_{ij} = w \delta_{i,j+1},
\end{align}
where $w > 0$ sets the gain. Then, $W$ has $N-1$ singular values identically equal to $w$, and therefore has rank $N-1$. As $(W^{k} u)_{j} = w^{k} \delta_{j,k+1}$, the chain can implement filters
\begin{align}
    f(t) = \v^{\top} e^{-(\mathbf{I}-\W) t}\u = e^{-t} \sum_{k=0}^{N-1} v_{k+1} \frac{(w t)^{k}}{k!}
\end{align}
where we absorb factors of $N$ and $\gamma$ as we are concerned with expressivity rather than optimization. 

Therefore, for a decaying target filter $f_{\star} = e^{-c_{\star} t}$, the problem of choosing $\v$ and $w$ to minimize the loss 
\begin{align}
    L = \int_{0}^{\infty} dt\, [f(t)-f_{\star}(t)]^2
\end{align}
reduces to the problem of how best to approximate $e^{-(c_{\star}-1) t}$ using polynomials of degree $N-1$ over $[0,\infty)$ with weight $e^{-2 t}$. The answer to this question is, of course, that we should choose $e^{t} f(t)$ to match the truncated expansion of $e^{t} f_{\star}(t)$ in Laguerre polynomials $L_{n}(2t)$. This perspective is reminiscent of the HiPPO framework for designing state-space models, which seeks to minimize the $L_2$ error over finite windows \cite{gu2020hippo}. 

Recall that $\int_{0}^{\infty} e^{-2t} L_{n}(2t) L_{m}(2t) \,dt = \frac{1}{2} \delta_{n,m}$. From the generating function of the Laguerre polynomials \cite{dlmf}, we have the expansion 
\begin{align}
    e^{(1-c_{\star}) t} = \frac{2}{c_{\star}+1} \sum_{n=0}^{\infty} \left(\frac{c_{\star}-1}{c_{\star}+1}\right)^{n} L_{n}(2t) , 
\end{align}
for which we provide a detailed derivation in the Supplemental Material \cite{supplement}. This sum is convergent in $L_2(e^{-2t})$ as 
\begin{align}
    \left(\frac{c_{\star}-1}{c_{\star}+1}\right)^{2} < 1
\end{align}
for all $c_{\star} > 0$. Thus, we must match the truncation of this sum to its first $N-1$ terms. Using the closed form
\begin{align}
    L_{n}(2t) = \sum_{k=0}^{n} \binom{n}{k} \frac{(-2t)^{k}}{k!},
\end{align}
this means that we should choose $w$ and $\v$ such that 
\begin{align}
    w^{k} v_{k+1}
    = \frac{2 (-2)^{k}}{c_{\star}+1} \sum_{n=k}^{N-1} \left(\frac{c_{\star}-1}{c_{\star}+1}\right)^{n} \binom{n}{k} . 
\end{align}
With this choice of weights, the error is given by the tail sum 
\begin{align}
    L &= \frac{2}{(c_{\star}+1)^2} \sum_{n=N}^{\infty} \left(\frac{c_{\star}-1}{c_{\star}+1}\right)^{2n} 
    \\
    &= \frac{1}{2 c_{\star}} \left(\frac{c_{\star}-1}{c_{\star}+1}\right)^{2N},
\end{align}
which decays geometrically in $N$. 

Therefore, for any given error tolerance, there exists a feedforward chain of some length that can solve the task within that tolerance. Concretely, if we demand that $L \leq \epsilon$ for some $\epsilon>0$, then there exists a satisfactory chain of length 
\begin{align}
    N \geq \frac{\log(2 c_{\star} \epsilon)}{\log[ (\frac{c_{\star}-1}{c_{\star}+1})^{2} ]} . 
\end{align}
Thus, for example, if we took $c_{\star} = 0.2$ as in Figure~\ref{fig:real_target}, then a chain of 50 neurons is sufficient to achieve a loss below $10^{-16}$. Unlike a low-rank network, this is not an exact solution (\textit{i.e.}, one that achieves loss exactly equal to zero), but it can be made arbitrarily close to one. This underscores the fact that the reason we do not find high-rank solutions is not because they cannot achieve good performance, but rather because of the inductive bias of gradient descent applied to this task. 

This construction differs from \citet{goldman2009feedback}'s classic construction of a chain-based approximate integrator, which seeks to uniformly approximate a target filter over compact intervals in time by Taylor expansion. Under that prescription, we choose weights $w^{k} v_{k+1} = (1-c_{\star})^{k}$ to match the Taylor expansion of $e^{(1-c_{\star}) t}$. In contrast, we seek to approximate in $L_{2}(e^{-2t})$, to match the infinite-horizon loss that we optimize elsewhere. Correspondingly, we can easily see that the Taylor series prescription leads to degraded approximation in this sense, as it will incur some error in the first $N-1$ modes, along with an identical tail sum.  

\section{Numerical methods and additional figures}

For our simulations, we implement the differentiable RNN rollout in JAX using \texttt{lax.scan()} \cite{jax2018github}. Since our theory is focused on gradient flow on the population loss, we define the loss function directly using the adjoints for the predictor $f(t)$, essentially averaging out over the random $x(t)$ sequences up front. This is legitimate to analyze the gradient flow at infinitesimal learning rate where minibatch fluctuations due to $\frac{\text{learning rate}}{\text{batchsize}}$ are negligible as well as discretization errors in the dynamics. We introduce a cutoff time $T$ which defines an interval $(0,T)$ over which we approximate the full RNN loss function. This interval is divided into $\mathcal N$ steps of size $\delta t \equiv \frac{T}{\mathcal N}$ and we use a small step-size discretization of the RNN dynamics with update step size $\delta t$. The loss function takes arguments $(\u, \v, \W)$ and computes $\mathcal L$ from the rollout
\begin{align}
    &\mathcal L \approx (\delta t) \sum_{ n = 1}^{\mathcal N} \left[ f(n\delta t) - f_\star(n \delta t) \right]^2  \nonumber
    \\
    &f(n \delta t) = \frac{1}{\gamma N} \v \cdot \u(n \delta t)  \nonumber
    \\
    &\u( (n+1)\delta t) = \u(n \delta t)  \nonumber
    \\
    &+ \delta t \times  \left[ - \mathbf I +   \frac{1}{\sqrt N} \W \right] \u(n \delta t) \nonumber
    \\
    &\u(0) = \u .
\end{align}
The target filter $f_\star(t) = e^{- c_\star t}$ for the real integrator task and $f_\star(t) = e^{-c_\star t} \cos(\omega_\star t)$ for the oscillator task. For most of our simulations $\delta t = 0.015$ and $\mathcal N = 2000$. We update the parameters using gradient descent on the loss $\mathcal L$ with learning rate $\eta$. We use learning rate $\eta = 0.01$ for most experiments. Complex eigenvalues are computed from the weights $\W$ using \texttt{numpy.linalg.eig()}. For experimental figures in the real integrator case, we obtain the numerical $c(s)$ value using the projection $\frac{\v_s^\top \W_s \u_s}{|\u_s||\v_s|}$. 

\begin{figure}[h!]
    \centering
    \includegraphics[width=3in]{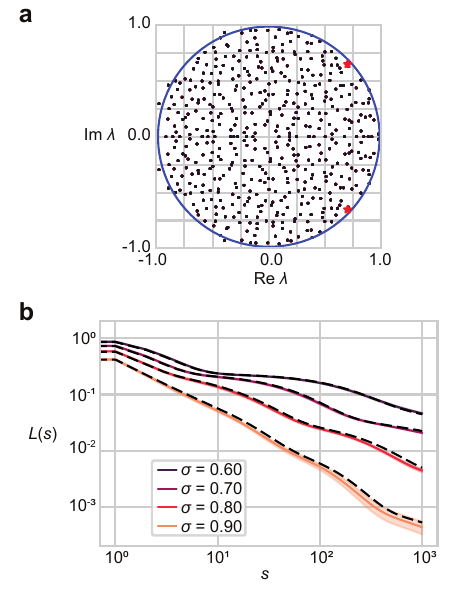}
    \caption{\textbf{Lazy learning of an oscillator.} \textbf{a}. Eigenvalue spectrum of a lazy network with $N=250$ neurons. The eigenvalues are approximately static over the course of learning. \textbf{b}. Loss dynamics in the lazy learning regime for $N = 4000$ networks $(c_{\star},\omega_{\star}) = (0.25,0.5)$. Dashed lines show the predictions of numerically solving for the lazy filter dynamics. }
    \label{fig:lazy_oscillator_supp}
\end{figure}

\begin{figure}[h!]
    \centering
    \includegraphics[width=0.75\linewidth]{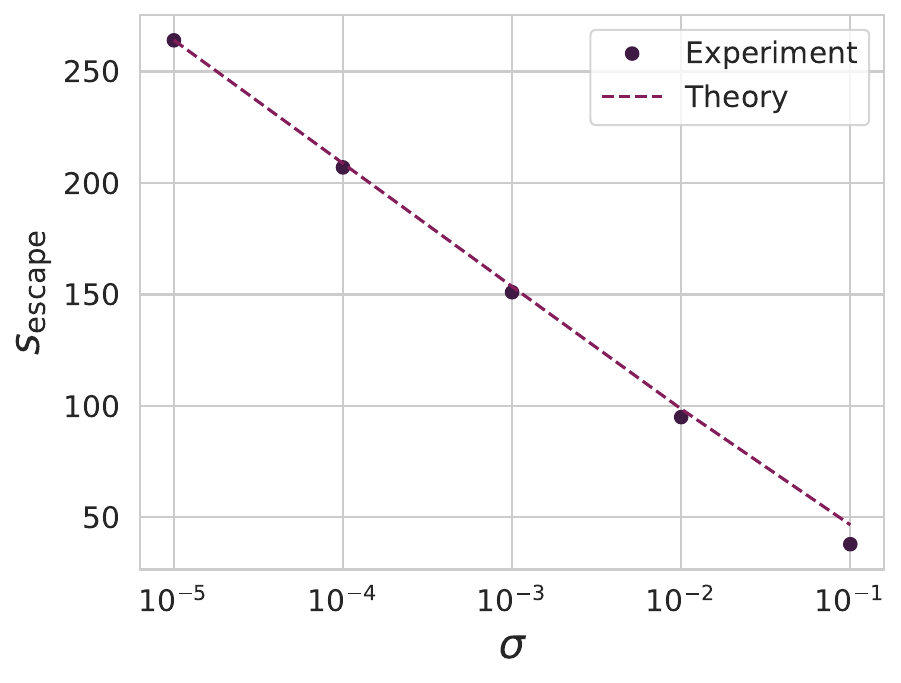}
    \caption{Escape time $s_{\text{escape}}$ for the real integration task predicted by our small $\sigma$ asymptotic theory compared to experiments at finite $\sigma$. }
    \label{fig:escape_vs_sigma_supp}
\end{figure}


\bibliography{refs}

\clearpage

\onecolumngrid

\begin{center}
    \textbf{SUPPLEMENTAL MATERIALS}

    \vspace{1em}
    In this supplemental document, we provide additional analysis of the lazy learning regime. This analysis uses a different approach than the direct feature-space analysis in Appendix B of the main text. Though it gives more detailed information about reservoir computing using the kernel $C$, many parts of the analysis cannot be extended easily to the NTK $K$. Moreover, we comment briefly on what principal component analysis would reveal about changes in dynamics in the very rich regime. 
\end{center}

\renewcommand{\thefigure}{S\arabic{figure}}
\setcounter{figure}{0}

\setcounter{section}{0}
\setcounter{figure}{0}

\renewcommand{\thesection}{S\arabic{section}}
\renewcommand{\thesubsection}{\thesection.\arabic{subsection}}
\renewcommand{\thesubsubsection}{\thesubsection.\arabic{subsubsection}}

\section{Preliminaries: the Mercer decomposition and kernel gradient flows}\label{app:mercer}

\subsection{The Mercer decomposition}

This approach starts by viewing $C$ (using it as a stand-in for a generic kernel) as the kernel of an integral operator $T_{C}$ acting on functions on $[0,\infty)$ \cite{berlinet2004rkhs}: 
\begin{align}
    [T_{C} f](t) = \int_{0}^{\infty} dt'\, C(t,t') f(t').
\end{align}
The trace of $C$ is
\begin{align}
    \int_{0}^{\infty} C(t,t) \, dt 
    &= \sigma^{2} \int_{0}^{\infty} e^{-2t} I_{0}(2 \sigma t) \, dt 
    \\
    &= \frac{\sigma^2}{2 \sqrt{1 -  \sigma^2}}
\end{align}
if $0 < \sigma < 1$; the kernel is otherwise not trace class because 
\begin{align}
    I_{0}(z) \sim \frac{e^{z}}{\sqrt{2 \pi z}}
\end{align}
as $z \to \infty$, which leads to a divergence \cite{dlmf}. This matches the expected threshold for stability of the linear ODE based on the spectral radius of $\W$. 

Presuming this condition holds, then $C$ admits a Mercer eigendecomposition 
\begin{align}
    C(t,t') = \sum_{n=0}^{\infty} \lambda_{n} \phi_{n}(t) \phi_{n}(t')
\end{align}
with orthonormal eigenvectors $\int_{0}^{\infty} dt\, \phi_{n}(t) \phi_{m}(t) = \delta_{nm}$ spanning $L_2$ and non-negative eigenvalues $\lambda_{n} \geq 0$. In terms of this eigendecomposition, the RKHS norm of a function $f$ is
\begin{align}
    \Vert f \Vert_{C}^2 = \sum_{n=0}^{\infty} \frac{|f_n|^2}{\lambda_{n}},
\end{align}
where we define the coefficients
\begin{align}
    f_{n} = \int_{0}^{\infty} dt\, \phi_{n}(t) f(t)
\end{align}
in terms of projections. Here, we write this expression quite formally: if $C$ has some zero modes---such that its eigenvectors with non-zero eigenvalue do not span $L_2$---then if $f_n$ is non-zero for those modes the RKHS norm diverges. 

In the main text, we use the alternative characterization of the RKHS norm as
\begin{align}
    \Vert f \Vert_{C}^{2} = \inf_{\{c_{n}\}}\left\{ \sum_{n=0}^{\infty} |c_{n}|^{2} \,\middle\vert\, \sum_{n=0}^{\infty} c_{n} \alpha_{n}(t) = f(t) \right\} 
\end{align}
where here $\alpha_{n}(t)$ are a generic (not necessarily orthogonal) feature realization of the kernel $C(t,t')$:
\begin{align}
    C(t,t') = \sum_{n=0}^{\infty} \alpha_{n}(t) \alpha_{n}(t'). 
\end{align}
This definition is clearly equivalent to that in terms of the Mercer decomposition, as we may use $\sqrt{\lambda_{n}} \phi_{n}(t)$ as features, in which case the coefficients are $c_{n} = f_{n}/\sqrt{\lambda_{n}}$. 

We will find that the Mercer decomposition of the reservoir computing kernel $C$ is an immediate consequence of known properties of Bessel functions. However, as we discuss in \S\ref{app:ntk_diagonalization}, this analysis regrettably does not extend easily to the NTK. There we show that the NTK $K(t,t')$ is tridiagonal in the basis of eigenvectors of $C(t,t')$, and is not of a form for which we are aware of a straightforward analytical diagonalization. In the main text, we circumvent this by using the feature-space formulation of the RKHS norm. 

\subsection{RKHS norm and parameter displacement in kernel gradient flows}\label{sec:gf_norm}

We now use the Mercer decomposition formally to show an important property of kernel gradient flows of the form studied in the main text. This is well-known, but we give a self-contained exposition. For the flow
\begin{align}
    \frac{\partial}{\partial s} f(t,s) = - \int_{0}^{\infty} C(t,t') [f(t',s) - f_{\star}(t')]\,dt'
\end{align}
starting from 
\begin{align}
    f(t,s=0) = 0,
\end{align}
the function $f(t,s)$ has finite RKHS norm $\Vert f(\cdot,s) \Vert_{C}$ at all fixed times $s < \infty$. However, if $f_{\star}$ lies outside the RKHS generated by $C$, then $\Vert f(\cdot,s) \Vert_{C}$ must diverge as $s \to \infty$. As a result, if we parameterize $f$ in terms of any set of features representing the kernel $C$, the function $f_{\star}$ cannot be learned with finite parameter norm. 

Assume $f_{\star}$ is in $L_2$, such that we can expand it as
\begin{align}
    f_{\star}(t) = \sum_{n=0}^{\infty} f_{\star,n} \phi_{n}(t).
\end{align}
Then, projecting the gradient flow into the eigenbasis, 
\begin{align}
    \frac{\partial}{\partial s} f_{n}(s) = - \lambda_{n} [f_{n}(s) - f_{\star,n}],
\end{align}
where
\begin{align}
    f(t,s) = \sum_{n=0}^{\infty} f_{n}(s) \phi_{n}(t). 
\end{align}
Then, we have the solution
\begin{align}
    f_{n}(s) = 
    \begin{cases}
        (1-e^{-\lambda_{n} s}) f_{\star,n} , & \lambda_{n} > 0 
        \\
        0 , & \lambda_{n} = 0,
    \end{cases}
\end{align}
as we assume the initial condition $f_{n}(s=0) = 0$. The RKHS norm of the solution is therefore
\begin{align}
    \Vert f(\cdot,s)\Vert_{C}^{2} = \sum_{\{n\mid \lambda_{n}>0\}} \frac{(1-e^{-\lambda_{n} s})^2}{\lambda_{n}} f_{\star,n}^2 .
\end{align}
Using the inequality $(1-e^{-x})^2 \leq 1-e^{-x} \leq x$, valid for all $x \geq 0$, we have 
\begin{align}
    \Vert f(\cdot,s)\Vert_{C}^{2} &\leq s \sum_{n=0}^{\infty} f_{\star,n}^{2} 
    = s  \Vert f_{\star} \Vert_{L_2}^{2}, 
\end{align}
which is always finite as we assume $f_{\star}$ is in $L_2$. However, as $s \to \infty$, $(1-e^{-\lambda_{n} s})^{2}$ tends to 1 from below for any fixed $n$ such that $\lambda_{n}>0$, which implies the monotone convergence (or divergence)
\begin{align}
    \lim_{s \to \infty} \Vert f(\cdot,s)\Vert_{C}^{2} = \sum_{\{n\mid \lambda_{n}>0\}} \frac{f_{\star,n}^2}{\lambda_{n}} = \Vert f_{\star}  \Vert_{C}^{2}.
\end{align}
This proves the claimed result: if $\Vert f_{\star}  \Vert_{C}^{2}$ is not finite, then $\Vert f(\cdot,s)\Vert_{C}^{2}$ must diverge. The exact rate of divergence will depend on the detailed scaling of the coefficients $f_{\star,n}^{2}$.

\section{Derivation of the Mercer decomposition using the Hardy-Hille formula}\label{sec:mercer}

Recall that
\begin{align}
    C(t,t') = \sigma^2 e^{-t - t'} I_{0}(2 \sqrt{ \sigma^2 t t'}).
\end{align}
Fortuitously, the Mercer decomposition of this kernel follows immediately from the Hardy-Hille formula for the Laguerre polynomials \cite{alsalam1964laguerre}: 
\begin{align}
    \sum_{n=0}^{\infty} L_{n}(x) L_{n}(y) q^{n} = \frac{1}{1-q} e^{-(x+y) q/(1-q) } I_{0}\left( \frac{2\sqrt{xyq}}{1-q}\right) . 
\end{align}
To obtain from this the Mercer decomposition of $C$, we put 
\begin{align}
    x = 2 c t, \quad y = 2 c t'
\end{align}
for an as-yet undetermined scale factor $c > 0$, and let 
\begin{align}
    \phi_{n}(t) = \sqrt{2c} e^{-c t} L_{n}(2 c t) .
\end{align}
The system of functions $\phi_{n}(t)$ are orthonormal with respect to Lebesgue measure:
\begin{align}
    \int_{0}^{\infty} dt\, \phi_{n}(t) \phi_{m}(t) = \delta_{nm},
\end{align}
and form a complete basis for $L^{2}([0,\infty))$. In terms of $\phi_{n}(t)$, we can write the Hardy-Hille formula as
\begin{align}
\begin{split}
    & \frac{1-q}{2c} \sigma^{2} \sum_{n=0}^{\infty} \phi_{n}(t) \phi_{n}(t') q^{n} \\&\quad = \sigma^{2} e^{-(t+t') 2 c (q/(1-q)+1/2)} I_{0}\left(\frac{4c\sqrt{ q tt'} }{1-q} \right) .
\end{split}
\end{align}
To match the desired expression for $C(t,t')$, we must have
\begin{align}
    2 c \left( \frac{q}{1-q} + \frac{1}{2} \right) = 1
    \quad \textrm{and} \quad 
    2 c \frac{\sqrt{q}}{1-q} = \sigma
\end{align}
which we can solve for
\begin{align}
    q = \frac{1-c}{1+c}
    \quad \mathrm{and} \quad 
    c = \sqrt{1-\sigma^2} .
\end{align}
Therefore, we have a Mercer decomposition 
\begin{align}
    C(t,t') = \sigma^{2} e^{-(t+t')} I_{0}(2 \sigma \sqrt{tt'}) = \sum_{n=0}^{\infty} \lambda_{n} \phi_{n}(t) \phi_{n}(t')
\end{align}
with orthonormal eigenvectors
\begin{align}
    \phi_{n}(t) &= \sqrt{2c} e^{-c t} L_{n}(2 c t) \\&=  \sqrt{2 \sqrt{1-\sigma^2}} e^{- \sqrt{1-\sigma^2} t} L_{n}(2 \sqrt{1-\sigma^2} t)
\end{align}
and corresponding eigenvalues
\begin{align}
    \lambda_{n} &= (1-c) q^n \\&= (1-\sqrt{1-\sigma^2}) \left(\frac{1- \sqrt{1-\sigma^2}}{1+\sqrt{1-\sigma^2}}\right)^{n}
\end{align}
As a sanity check, we observe that
\begin{align}
    \sum_{n=0}^{\infty} \lambda_{n} = \frac{\sigma^2}{2c} = \frac{\sigma^2}{2\sqrt{1-\sigma^2}},
\end{align}
as we found from direct computation of the integral, where the series is summable for $0 < \sigma < 1$. 

\begin{figure*}[t]
    \centering
    \includegraphics[width=7in]{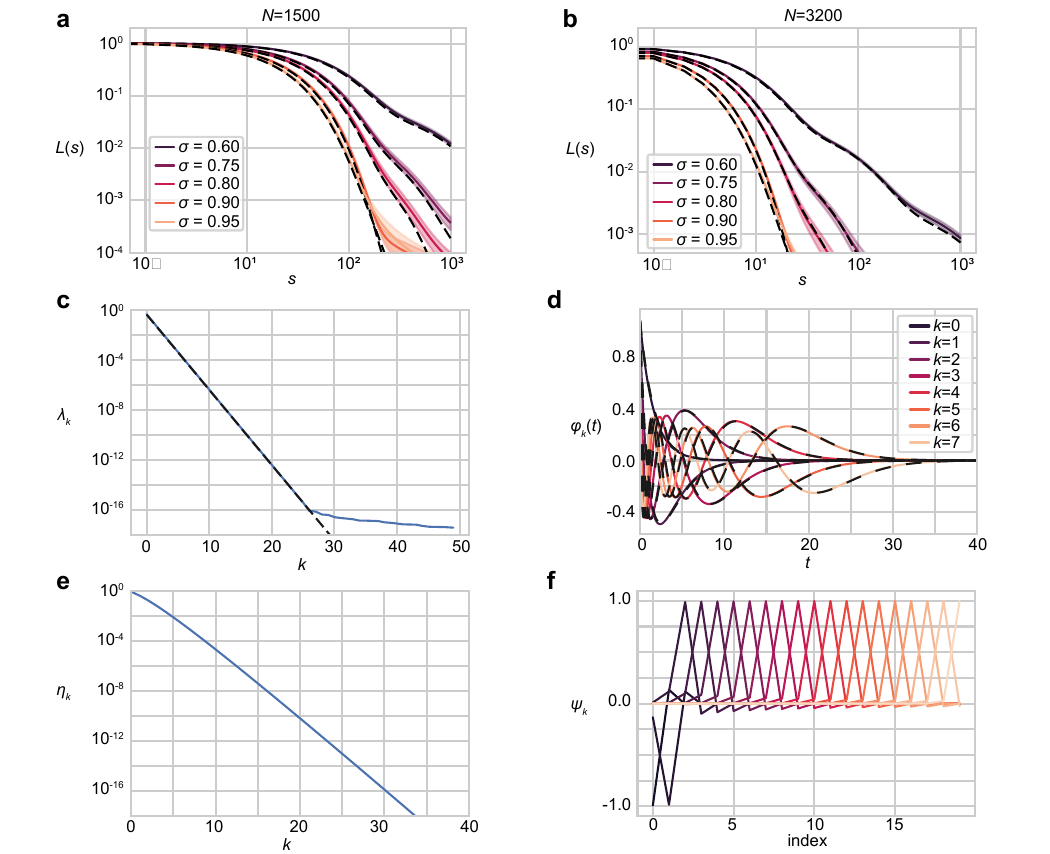}
    \caption{\textbf{Additional figures on lazy integrator learning and kernel eigendecompositions.} \textbf{a}. Lazy training of networks of size $N = 1500$ with varying initialization variance $\sigma^2$ on a task with $c_{\star} = 0.5$.  Black dashed lines show theory predictions resulting from numerical solution of the lazy learning dynamics with the finite-size empirical NTK. Compare to Figure~2c's $N=4000$ networks. \textbf{b}. As in \textbf{a}, but for networks of size $N = 3200$. \textbf{c}. Eigenvalue spectrum of the DMFT autocorrelation $C(t,t')$ for $\sigma = 0.8$. Blue line shows result of numerical diagonalization of $C(t,t')$ sampled on a grid with temporal resolution $\Delta t = 0.1$ up to $t = 100$ using double-precision floating point arithmetic. Black dashed line shows the analytical result from Appendix \ref{app:mercer}. Discrepancies emerge around the working precision $2^{-52} \sim 10^{-16}$. \textbf{d}. The first eight eigenvectors $\phi_{k}(t)$ of $C(t,t')$, in the same setting as in \textbf{c}. Black dashed lines show the eigenvectors obtained analytically in Appendix \ref{app:mercer}. \textbf{e}. Estimated eigenvalues $\eta_k$ of the NTK based on numerical diagonalization of its tridiagonal matrix form in the basis of eigenvectors of $C(t,t')$ (see Appendix \ref{app:mercer}). The numerics are based on a truncation of the infinite tridiagonal matrix to a $100 \times 100$ array. \textbf{f}. As in \textbf{e}, but showing the eigenvectors $\psi_k$ of the NTK in the autocorrelation eigenvector basis shown in \textbf{d}. }
    \label{fig:lazy_integrator_supp}
\end{figure*}

In Figure \ref{fig:lazy_integrator_supp}c-d, we compare this analytical result to numerical diagonalization of $C(t,t')$ sampled on a discrete grid. We see excellent agreement of the first few eigenvectors and eigenvalues, but discrepancies emerge beyond the first 25 or so eigenvalues as their exponential decay means they quickly fall below the working precision of double-precision floating point arithmetic. 

\section{Expansion of an exponential decay in the eigenbasis of \texorpdfstring{$C$}{C}}\label{app:exponential_decay}

Let us now expand an exponential with a general time constant $e^{-c_{\star} t}$ in this basis. We do so starting from the generating function for the Laguerre polynomials \cite{dlmf}, 
\begin{align}
    \sum_{n=0}^{\infty} r^{n} L_{n}(x) = \frac{1}{1-r} e^{-r x/(1-r)} 
\end{align}
for $|r| < 1$. Putting $x = 2 c t$ and multiplying by $\sqrt{2c} e^{-c t}$, we have
\begin{align}
    \sum_{n=0}^{\infty} r^{n} \phi_{n}(t) = \frac{1}{1-r} \sqrt{2c} e^{-2 c \left(1/2 + r/(1-r)\right) t}.
\end{align}
We therefore see that we must have
\begin{align}
    2 c \left(\frac{1}{2} + \frac{r}{1-r} \right) = c_{\star}
\end{align}
which we can solve to obtain
\begin{align}
    r = \frac{c_{\star} - c}{c_{\star} + c}
\end{align}
Re-arranging, we can see that
\begin{align}
    e^{-c_{\star} t} = \frac{1-r}{\sqrt{2c}} \sum_{n=0}^{\infty} r^{n} \phi_{n}(t) .
\end{align}
As a sanity check, noting that we always have $r^2<1$, computing the $L_2$ norm from this expansion gives 
\begin{align}
    \frac{(1-r)^2}{2c} \sum_{n=0}^{\infty} r^{2n} = \frac{(1-r)^2}{2c} \frac{1}{1-r^2} = \frac{1}{2c_{\star}}
\end{align}
as expected. Now we consider the RKHS norm
\begin{align}
     \Vert e^{-c_{\star} t} \Vert_{C}^{2} &= \frac{(1-r)^2}{2c} \sum_{n=0}^{\infty} \frac{1}{\lambda_{n}} r^{2n} 
     \\
     &= \frac{(1-r)^2}{1-q} \frac{1}{\sigma^2} \sum_{n=0}^{\infty}  \left(\frac{r^2}{q}\right)^{n} .
\end{align}
Therefore, for the RKHS norm to be finite, we must have
\begin{align}
    1 > \frac{r^2}{q} = \left(\frac{c_{\star}-c}{c_{\star}+c}\right)^{2} \frac{1+c}{1-c} . 
\end{align}
This ratio is a non-monotonic function of $c_{\star}$ for each $c$: as $c_{\star}$ increases from zero, it decreases towards a minimum at $c_{\star} = c$, before increasing again. It is less than unity for
\begin{align}
    1 - \sqrt{1-c^2} < c_{\star} < 1 + \sqrt{1-c^2}
\end{align}
which, as $c = \sqrt{1-\sigma^2}$, translates to
\begin{align}
    1 - \sigma < c_{\star} < 1 + \sigma. 
\end{align}
This corresponds precisely to the spectral edges of the initial dynamics matrix. In other words, an exponential decay is in-RKHS---and therefore lazily learnable---iff its decay timescale lies within the spectrum of timescales of the reservoir. If this condition is met, we can sum the geometric series to obtain the explicit expression
\begin{align}
     \Vert e^{-c_{\star} t} \Vert_{C}^{2} 
     = \frac{1}{\sigma^2 - (1-c_{\star})^2}. 
\end{align}
This agrees with the result we found in the main text. 

If $c_{\star}$ is outside the spectral edges, we can compute how quickly the RKHS norm diverges as a function of training time $s$. For convenience, we will ignore the critical case in which $c_{\star}$ lands right on the edge. Then, substituting in the particular expressions for $\lambda_{n}$ and $f_{\star,n}^{2}$ we have into the expression for the RKHS norm $\Vert f(\cdot,s)\Vert_{C}^{2}$ we found in Section~\ref{sec:gf_norm}, we obtain
\begin{align}
    \Vert f(\cdot,s)\Vert_{C}^{2} = \frac{(1-r)^2}{2c(1-c)}  \sum_{n=0}^{\infty} (1-e^{-(1-c) q^{n} s})^2 \left(\frac{r^2}{q}\right)^{n}, 
\end{align}
where
\begin{align}
    r = \frac{c_{\star}-c}{c_{\star}+c}
    \quad \mathrm{and} \quad
    q = \frac{1-c}{1+c}
\end{align}
are given in terms of $c = \sqrt{1-\sigma^2}$. In the regime of interest, $r^2>q$, and we know that the sum diverges as $s \to \infty$. To determine the scaling, we observe that the prefactor multiplying $(r^2/q)^n$ is exponentially small unless $(1-c) q^{n} s$ is large. As $0<q<1$, for a given $s$ this means that low-lying modes $n \leq M$ dominate the sum, where the cutoff $M$ is determined by the condition that $(1-c) s q^{M} \sim 1$, \textit{i.e.}, we have that $M$ scales like
\begin{align}
    M \sim \frac{\log[(1-c)s]}{\log(1/q)}.
\end{align}
Therefore, the dominant behavior of the sum is
\begin{align}
    \sum_{n=0}^{\infty} (1-e^{-(1-c) q^{n} s})^2 \left(\frac{r^2}{q}\right)^{n}
    \sim \sum_{n=0}^{M}  (1-e^{-(1-c) q^{n} s})^2 \left(\frac{r^2}{q}\right)^{n}
    \sim \sum_{n=0}^{M} \left(\frac{r^2}{q}\right)^{n}
    \sim \left(\frac{r^2}{q}\right)^{M} . 
\end{align}
Substituting in the scaling of $M$ and simplifying, this leads to 
\begin{align}
    \Vert f(\cdot,s)\Vert_{C}^{2}
    \sim \left(\frac{r^2}{q}\right)^{M}
    \sim s^{\gamma},
\end{align}
where 
\begin{align}
    \gamma = \frac{\log(r^2/q)}{\log(1/q)} = \frac{\log[(\frac{c_{\star}-c}{c_{\star}+c})^2 \frac{1+c}{1-c}]}{\log\frac{1+c}{1-c}} . 
\end{align}
Therefore, we find that the RKHS norm diverges as a power law in time. However, the exponent of this divergence is not very large: by inspection we see that it is maximized by taking $c_{\star} \downarrow 0$, where $\gamma \uparrow 1$. Thus, if we stop the gradient flow at a fixed, finite time, the squared RKHS norm can grow at most linearly with that early-stopping time. This is consistent with the generic bound $\Vert f(\cdot,s)\Vert_{C}^{2} \leq s \Vert f_{\star} \Vert_{L_2}^{2}$ which we obtained in Section~\ref{sec:gf_norm}. If $c_{\star}$ is just outside the interval $[1-\sigma,1+\sigma]$, then the exponent $\gamma$ will be very close to zero, and the divergence will be quite slow. Indeed, right at the spectral edge, we expect the norm to diverge logarithmically.

\section{Can we extend this to the NTK?}\label{app:ntk_diagonalization}

We now want to consider the NTK
\begin{align}
    K(t_{1},t_{2}) = 2 C(t_{1},t_{2}) + C^{\ast 2}(t_{1},t_{2}), 
\end{align}
where $C^{\ast 2}(t_{1},t_{2})$ is the convolution
\begin{align}
    C^{\ast 2}(t_{1},t_{2}) = \int_{0}^{t_{1}} dt_{3}\, \int_{0}^{t_{2}} dt_{4}\, C(t_{3},t_{4})  C(t_{1}-t_{3}, t_{2}-t_{4}).
\end{align}
Substituting in the Mercer decomposition of $C$, we have
\begin{align}
\begin{split}
    C^{\ast 2}(t_{1},t_{2}) = \sum_{n,m=0}^{\infty} \lambda_{m} \lambda_{n} &\int_{0}^{t_{1}} dt_{3}\,\phi_{n}(t_{3})\phi_{m}(t_{1}-t_{3}) \\\times & \int_{0}^{t_{2}} dt_{4}\,\phi_{n}(t_{4})\phi_{m}(t_{2}-t_{4}). 
\end{split}
\end{align}
Using the fact that
\begin{align}
    \phi_{n}(t) = \sqrt{2 c} e^{-c t} L_{n}(2 c t),
\end{align}
we have
\begin{align}
    \int_{0}^{t_{1}} dt_{3}\,\phi_{n}(t_{3})\phi_{m}(t_{1}-t_{3})
    = e^{-x/2} \int_{0}^{x} dy\, L_{n}(y) L_{m}(x-y),
\end{align}
where we let $x = 2 c t_{1}$ and $y = 2 c t_{3}$. Using the identity
\begin{align}
    \int_{0}^{x} dy\, L_{n}(y) L_{m}(x-y) = L_{n+m}(x) - L_{n+m+1}(x)
\end{align}
from DLMF 18.17.2 \cite{dlmf}, we thus find that
\begin{align}
    \int_{0}^{t_{1}} dt_{3}\,\phi_{n}(t_{3})\phi_{m}(t_{1}-t_{3})
    &= e^{-x/2} L_{n+m}(x) - e^{-x/2} L_{n+m+1}(x)
    \\
    &= (2c)^{-1/2} [\phi_{n+m}(t_1) - \phi_{n+m+1}(t_1)]. 
\end{align}
Therefore, we have
\begin{align}
    C^{\ast 2}(t_{1},t_{2}) = \frac{1}{2c} \sum_{n,m=0}^{\infty} \lambda_{m} \lambda_{n} [\phi_{n+m}(t_1) - \phi_{n+m+1}(t_1)] [\phi_{n+m}(t_2) - \phi_{n+m+1}(t_2)]. 
\end{align}
It follows that the matrix elements of this convolution in the basis of eigenvectors of $C$ are
\begin{align}
    [C^{\ast 2} ]_{nm} =\frac{1}{2c} \sum_{k,l=0}^{\infty} \lambda_{k} \lambda_{l} [\delta_{k+l,n} - \delta_{k+l+1,n}] [\delta_{k+l,m} - \delta_{k+l+1,m}] . 
\end{align}
Substituting in $\lambda_{k} = (1-c) q^{k}$, and evaluating the sums, we have
\begin{align}
     [C^{\ast 2} ]_{nm} = \frac{(1-c)^2}{2c}\bigg( [n q^{n-1} + (n+1) q^{n}] \delta_{n,m} - (m+1) q^{m} \delta_{n,m+1} - (n+1) q^{n} \delta_{m,n+1} \bigg) .
\end{align}
Thus, the NTK is tridiagonal in the basis of eigenvectors of the DMFT autocorrelation, with matrix elements
\begin{align}
    [K]_{nm} =
     \left( 2 (1-c) q^{n} + \frac{(1-c)^2}{2c} [n q^{n-1} + (n+1) q^{n}] \right) \delta_{n,m}  - \frac{(1-c)^2}{2c} \bigg((m+1) q^{m} \delta_{n,m+1} + (n+1) q^{n} \delta_{m,n+1} \bigg).
\end{align}
We have as yet not succeed in analytically determining the eigenvalues and eigenvectors of this infinite symmetric tridiagonal matrix. However, as its entries decay exponentially, we can estimate its eigenvectors and eigenvalues numerically by truncating at some fixed size and using a solver for the tridiagonal symmetric eigenproblem (we use SciPy's \texttt{eigh\_tridiagonal}). We show the results of this numerical estimation in Figure \ref{fig:lazy_integrator_supp}e-f. The eigenvalues decay exponentially, and the eigenvectors are approximately localized. Though a full analysis of these properties are beyond the scope of the present work, they pose an interesting mathematical problem. 

\section{Principal component analysis of activity in the rich regime}

Our work mostly views the dynamics of the linear RNN in weight space and at the level of the induced filter $f(t)$. One might wonder whether further insight can be gained through additional analyses of the activity $\h$. In particular, as principal component analysis (PCA) is often applied to study neural dynamics in task-trained RNNs \cite{driscoll2024flexible,qian2024partial,qian2026discovering}, one might naturally wonder whether it reveals anything interesting about how dynamics change through learning. 

However, in our setting, PCA both before and after learning in the very rich regime $\sigma \ll 1$ would reveal activity concentrated along a one-dimensional subspace. This is because activity at initialization is concentrated along the input direction $\u$ when $\sigma \ll 1$, while after rich learning it is concentrated along the aligned direction. 

To see this, observe that the noise-driven linear RNN with fixed weight matrix is a multidimensional Ornstein-Uhlenbeck process. Its stationary covariance $\bm{\Sigma} = \lim_{t \to \infty} \mathbb{E}[\mathbf{h}(t) \mathbf{h}(t)^{\top}]$ is given by \cite{gardiner1985handbook}
\begin{align}
    \bm{\Sigma} = \int_{0}^{\infty} e^{-(\mathbf{I}-\frac{1}{\sqrt{N}} \mathbf{W}) t} \u \u^{\top} e^{-(\mathbf{I}-\frac{1}{\sqrt{N}} \mathbf{W})^{\top} t} \,dt. 
\end{align}

Suppose we are at initialization. For very small $\sigma$, we can neglect the factors of $\mathbf{W}$ in the exponent. To make this precise, write $\mathbf{W} = \sigma \mathbf{G}$, where $\mathbf{G}$ is a standard Ginibre matrix, \textit{i.e.}, one with i.i.d. $\mathcal{N}(0,1)$ elements. Then, we have the formal expansion 
\begin{align}
    \bm{\Sigma} 
    &= \int_{0}^{\infty} e^{\frac{\sigma}{\sqrt{N}}\mathbf{G} t} \u \u^{\top} e^{\frac{\sigma}{\sqrt{N}} \mathbf{G}^{\top} t} e^{-2t}\,dt
    \\
    &= \frac{1}{2} \sum_{n,m=0}^{\infty} \frac{(n+m)!}{n!\, m!} \left(\frac{\sigma}{2\sqrt{N}}\right)^{n+m} \mathbf{G}^{n} \u \u^{\top} (\mathbf{G}^{\top})^{m}
    \\
    &= \frac{1}{2} \sum_{n=0}^{\infty} \left(\frac{\sigma}{2\sqrt{N}}\right)^{n}  \sum_{m=0}^{n} \binom{n}{m} \mathbf{G}^{m} \u \u^{\top} (\mathbf{G}^{\top})^{n-m} 
\end{align}
as $\int_{0}^{\infty} t^{n} e^{-2t }\, dt = \frac{n!}{2^{n+1}}$. Here, the interchange of the integral over $t$ with the infinite sums is justified with high probability for any $\sigma < 1$ as $\max_{i} \Re \lambda_{i}(\frac{1}{\sqrt{N}} \mathbf{G} ) \leq 1$ with high probability. Moreover, $\Vert \frac{1}{\sqrt{N}} \mathbf{G} \Vert_{op} \leq 2$, so the scale of a given term in the sum is governed by the $\sigma^{n}$ prefactor. Thus, at small $\sigma$, the first term will dominate:
\begin{align}
    \bm{\Sigma} = \frac{1}{2} \u \u^{\top} + \mathcal{O}(\sigma^3),
\end{align}
where we note that $\u \u^{\top}$ is itself $\mathcal{O}(\sigma^2)$. This means that the initial dynamics will appear approximately one-dimensional under PCA. 

Now consider the endpoint of learning in the very rich regime, with a target filter $e^{-c_{\star} t}$. In this case, we have $\frac{1}{\sqrt{N}} \mathbf{W} \approx (1-c_{\star}) \hat{\u} \hat{\u}^{\top}$ and $\u = U \hat{\u}$ for a unit vector $\hat{\u}$ and a constant $U>0$, determined by the alignment phase. Then, $e^{\frac{1}{\sqrt{N}} \mathbf{W} t} \approx e^{(1-c_{\star}) t} \hat{\u} \hat{\u}^{\top}$, so 
\begin{align}
    \bm{\Sigma} \approx \frac{U^2}{2 c_{\star}} \hat{\u} \hat{\u}^{\top} . 
\end{align}
Thus, once again, the dynamics look approximately one-dimensional under PCA. 

At large $N$, we can in fact say something sharper about the eigenvalues of $\bm{\Sigma}$ at initialization. Let $\lambda_{j}(\bm{\Sigma}/N)$ be the $j$-th ordered eigenvalue of $\bm{\Sigma}/N$, with $\lambda_{0}$ being the largest eigenvalue. Then, for each fixed $n \geq 0$, we have that 
\begin{align}
    \lambda_{n}\left(\frac{\bm{\Sigma}}{N}\right) \to (1-\sqrt{1-\sigma^2}) \left(\frac{1-\sqrt{1-\sigma^2}}{1+\sqrt{1-\sigma^2}}\right)^{n}.
\end{align}
That is, the macroscopic eigenvalues of $\frac{1}{N} \bm{\Sigma}$ are precisely the Mercer eigenvalues of the operator $C$ that we studied in Section~\ref{sec:mercer}. Heuristically, this follows from a temporal discretization: if we define the matrix $\mathbf{H}$ with elements $H_{tj} = \frac{1}{\sqrt{N}} [e^{-(\mathbf{I} - \frac{1}{\sqrt{N}} \W) t} \u]_{j}$ over some discretized grid of times, then we can write $\frac{1}{N} \bm{\Sigma} = \mathbf{H}^{\top} \mathbf{H}$ and $\mathbf{C} = \mathbf{H} \mathbf{H}^{\top}$, which clearly have the same non-zero eigenvalues. Then, it becomes an issue of ordering of limits in $N$ and continuous time, taking into account somehow the fact that the continuous time limit must also extend over the full semi-infinite interval. 

To avoid the potential subtleties arising there, we take an alternate route. From above, we have a series expansion for $\bm{\Sigma}$. Defining the vectors
\begin{align}
    \mathbf{p}_{n} = \frac{1}{N^{(n+1)/2} \sigma} \mathbf{G}^{n} \u 
\end{align}
and the matrix
\begin{align}
    A_{mn} = \frac{(n+m)!}{n!\,m!} \frac{\sigma^{n+m+2}}{2^{n+m+1}} ,
\end{align}
we can re-write the expansion we found before as
\begin{align}
    \frac{1}{N} \bm{\Sigma} = \sum_{n,m=0}^{\infty} A_{mn} \mathbf{p}_{n}\mathbf{p}_{m}^{\top}. 
\end{align}

Now, for fixed $n,m \geq 0$, we have that
\begin{align}
    \mathbf{p}_{n}^{\top} \mathbf{p}_{m} \to \delta_{nm}
\end{align}
in probability as $N \to \infty$. This follows from Wick's theorem: the dominant contribution to $\mathbb{E}[\mathbf{p}_{n}^{\top}\mathbf{p}_{m}]$ comes from pairings of $\mathbf{G}$ with $\mathbf{G}^{\top}$. If $n \neq m$, then fewer than $(n+m)/2$ such parings can be made, and the expectation vanishes as $N \to \infty$. Similarly, the dominant contribution to the second moment comes from pairs of such pairings, but that contribution is subtracted off by the copies of the mean appearing in the variance. 

Our task is therefore to show that the eigenvalues of the infinite symmetric matrix $\mathbf{A}$ coincide with the Mercer eigenvalues of $C$. This follows via a more rigorous version of the ``$\frac{1}{N} \bm{\Sigma} = \mathbf{H}^{\top} \mathbf{H}$; $\mathbf{C} = \mathbf{H} \mathbf{H}^{\top}$'' heuristic above, now at the level of operators. Using the series expansion of the Bessel function as in Appendix B of the main text, we can write 
\begin{align}
    C(t,t') = \sum_{n=0}^{\infty} \alpha_{n}(t) \alpha_{n}(t') 
\end{align}
in terms of features 
\begin{align}
    \alpha_{n}(t) = \frac{1}{n!} e^{-t} \sigma^{n+1} t^{n} . 
\end{align}
But, we have
\begin{align}
    \int_{0}^{\infty} \alpha_{n}(t) \alpha_{m}(t)\,dt 
    &= \frac{\sigma^{n+m+2}}{n!\,m!} \int_{0}^{\infty} t^{n+m} e^{-2t}\,dt
    \\
    &= \frac{(n+m)!}{n!\,m!} \frac{\sigma^{n+m+2}}{2^{n+m+1}}
    \\
    &= A_{nm} . 
\end{align}
That is, $\mathbf{A}$ is the Gram matrix of the features $\alpha_{n}(t)$. As a result, it must have the same non-zero spectrum as the Mercer eigenvalues of $C(t,t')$, which completes the argument. 

\end{document}